\documentclass[aps,floats,twocolumn,epsf,prb,superscriptaddress,showpacs]{revtex4-1}
\usepackage{graphicx}
\usepackage{epstopdf}
\usepackage{amsmath}
\usepackage{amssymb}
\usepackage{booktabs}
\usepackage[colorlinks=true,urlcolor=blue,citecolor=blue,linkcolor=blue,breaklinks=true]{hyperref}
\usepackage{color}
\usepackage{bm}
\usepackage{mathpazo}
\usepackage{dsfont}
\usepackage[dvipsnames]{xcolor}
\usepackage{subcaption}
\usepackage{caption}

\usepackage{orcidlink} % orchidlink.sty needs to be present/uploaded.

\newcommand{\braket}[1]{\langle #1 \rangle}

\newcommand{\kh}[1]{{\color{ForestGreen}#1}}

 \DeclareCaptionJustification{links}{\raggedright}
\newcommand{\hhat}{\hat{\mathbf{h}}}

\begin{document}

\preprint{APS/123-QED}

\title{Weyl points and spin-orbit coupling in copper-substituted lead phosphate apatite}

%\author{M. Braß}
%\author{Liang Si}%
%\altaffiliation[Also at ]{School of Physics, Northwest University, Xi'an 710127, China}
%\author{K. Held}
%\affiliation{Institute of Solid State Physics, TU Vienna}

\author{Martin Bra\ss\,\orcidlink{0000-0002-4347-6987}}
\affiliation{Institute of Solid State Physics, TU Wien, 1040 Vienna, Austria}

\author{Liang Si\,\orcidlink{0000-0003-4709-6882}}
\affiliation{School of Physics, Northwest University, Xi'an 710127, China}
%\affiliation{Institute of Solid State Physics, TU Wien, 1040 Vienna, Austria}

\author{Karten Held\,\orcidlink{0000-0001-5984-8549}}
\affiliation{Institute of Solid State Physics, TU Wien, 1040 Vienna, Austria}

\date{\today}

\begin{abstract}
 We study the impact of spin-orbit coupling on the topological band-properties of copper-substituted lead phosphate apatite using a combination of group-theoretical analysis and full-relativistic density-functional theory calculations. We characterize Weyl points at time-reversal invariant momenta and find that a band-inversion due to spin-orbit coupling leads to additional Weyl points close to the Fermi-edge at general momenta. To determine the position of the altogether 66 Weyl points in the Brilouin-zone, we develop an algorithm that follows a Berry-curvature-derived vector field to its monopole: the Weyl point. The emerging surface Fermi-arcs and their spin-polarization reveal avoided crossings and a Fermi-loop detached from the Weyl points.
\end{abstract}

\maketitle

\section{\label{sec:Introduction}Introduction}
%\kh{[KH: did not mark changes in the Intro in color, as it is very much overhauled]}
Recent pronouncements of room temperature superconductivity \cite{lee2023,lee2023a,lee2023b} have put 
 copper substituted lead apatite Pb$_9$Cu(PO$_4$)$_6$O into the focus of solid state research.   Subsequent work has shown that -- without further doping --  Pb$_{10-x}$Cu$_x$(PO$_4$)$_6$O is a Mott or charge transfer insulator for all $x$~\cite{si2023,Si2023b,Si2023b,Korotin2023,yue2023correlated,liu2023symmetry,georgescu2023,kumar2023absence,puphal2023single,jiang2023pb,liu2023,wang2023ferromagnetic}, and that the observed conductivity jumps likely originate from residual Cu$_2$S \cite{liu2023,Zhu2023,Jain2023}. 

 Even without superconductivity,  Pb$_9$Cu(PO$_4$)$_6$O  is interesting in its own right. Its bandstructure in density fucntional theory (DFT) \cite{si2023,griffin2023,lai2024,cabezas-escares2023} exhibits two almost flat bands which cross the Fermi-edge and are formed by the Cu $d_{xz}/d_{yz}$ orbitals. The symmetry properties of the compound imply that these bands contain Weyl points \cite{zhou2023,hirschmann2023} which may impact thermal as well as electro-magnetic transport properties and give rise to topologically protected surface states.  
%
% Liang,m I would leave thuis dioscussion on SC out, most likely it is not SC.
%\ls{However, a deep understanding of the electronic and putative pairing force of electrons in LK-99 is still lacking.
%The pairing mechanism driving the putative superconductivity remains highly controversial since the very first experimental reports \cite{}. Density-functional theory \cite{} and dynamical mean-field theory calculations \cite{} have been performed to investigate its similarities and differences between two-dimensional cuprate superconductors \cite{}. Different proposals have been proposed, such as flat bands \cite{}, magnetic spin \cite{}, charge fluctuations \cite{} and correlations in Cu-3$d$ orbitals have been proposed. The possible inhomogenous distribution of Cu ions may lead to further complicated electronic and magnetic states in the grown process of LK-99. Additionally, by employing different synthesis techniques \cite{}, different ground states \cite{} had been determined to explain the nature of LK-99.
%
Despite its importance and topicality, Weyl points in  Pb$_9$Cu(PO$_4$)$_6$O  aka LK-99 have not been deeply investigated hitherto.
%his hence raises the opening question of whether the poissible Weyl points are able to enhance/destroy the putative SC or even induce another superconductive resource.
%}

Here, we construct a tight-binding model from DFT calculations of the electronic band-structure in section \ref{sec:DFT}. We study the topological properties of this compound with a group-theoretical analysis to characterize all symmetry protected Weyl points at time-reversal invariant momenta in the presence and absence of spin-orbit coupling (SOC) in section \ref{sec:Symmetry}. In combination with the band-structure calculations this uncovers the emergence of additional Weyl points at general momenta  due to the influence of SOC. To detect these Weyl points automatically we present a novel algorithm in section \ref{sec:Algorithm}. With SOC, there are additional Weyl points close to the Fermi surface resulting in surface Fermi-arcs as described in section \ref{sec:Surface}.

Before starting, let us put some caveats here: (i) We consider the lowest energy structure (for a single unit cell) of Pb$_9$Cu(PO$_4$)$_6$O, as shown in Fig.~\ref{fig:structure}. This has a  \emph{P3} (no.~143) space group, but other structures (other orientations of the ``extra" or channel O and of the Cu) are only $\sim$6\,meV per unit cell different in energy\cite{si2023}. This means that observing a single crystal instead of a disordered compound requires temperatures well below 6\,meV (70\,K) or under pressure $>$ 73\,GPa according to DFT calculations \cite{yang2023ab}.

(ii) Since Pb$_{10-x}$Cu$_x$(PO$_4$)$_6$O is insulating, a slight electron or hole doping is required to obtain the Weyl points studied here on the DFT level. Such an electron or hole doping is not possible by changing $x$ as Cu and Pb are both $2+$. Instead O excess or deficiency, substituting P by S or other means that change the valence on the Cu sites is needed.
With such a doping, a quasiparticle peak will emerge at the Fermi level which is a renormalized (more narrow) 
version of the electronic structure analyzed in the present paper.
%\section{\label{sec:Methods}Methods and Results}

\section{\label{sec:DFT}Density Functional Theory}
\begin{figure*}
    \centering
    \begin{subfigure}{0.4\textwidth}
        \includegraphics[width=\textwidth]{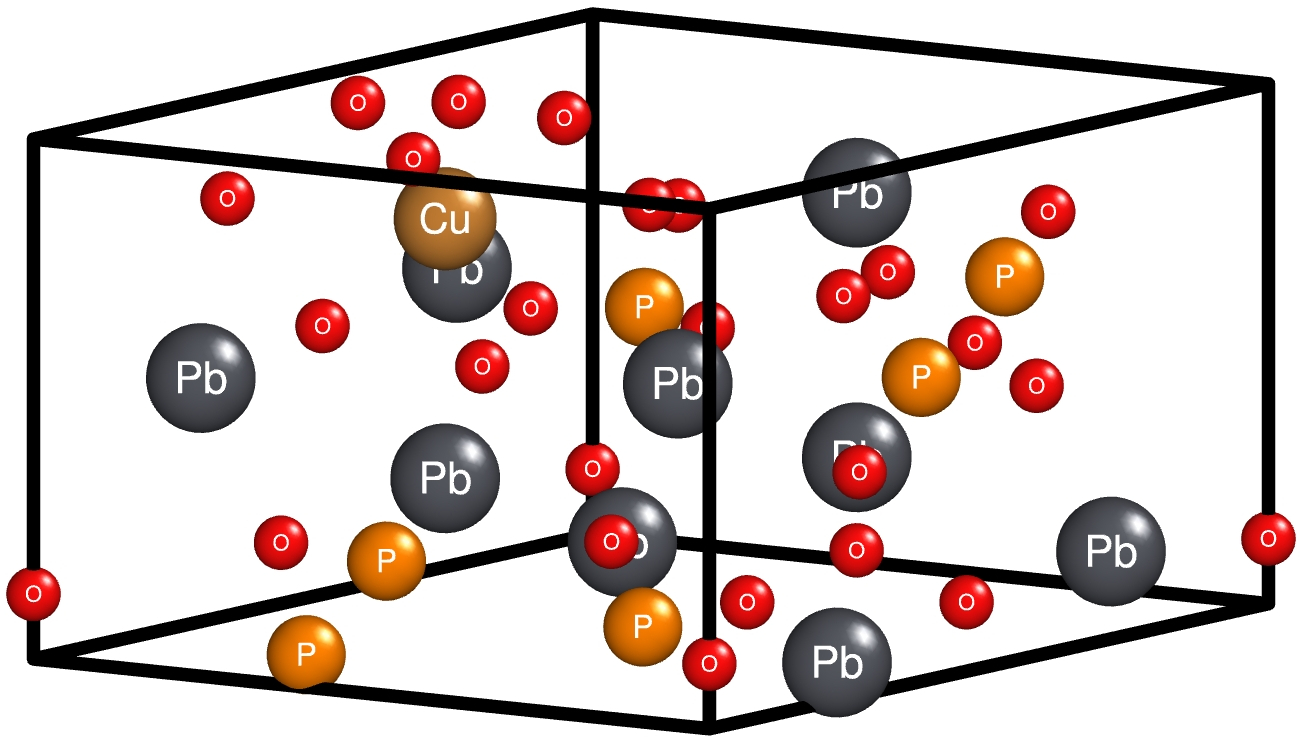}
        \caption{\label{fig:structure}}
    \end{subfigure}
    \begin{subfigure}{0.59\textwidth}
        \includegraphics[width=\textwidth]{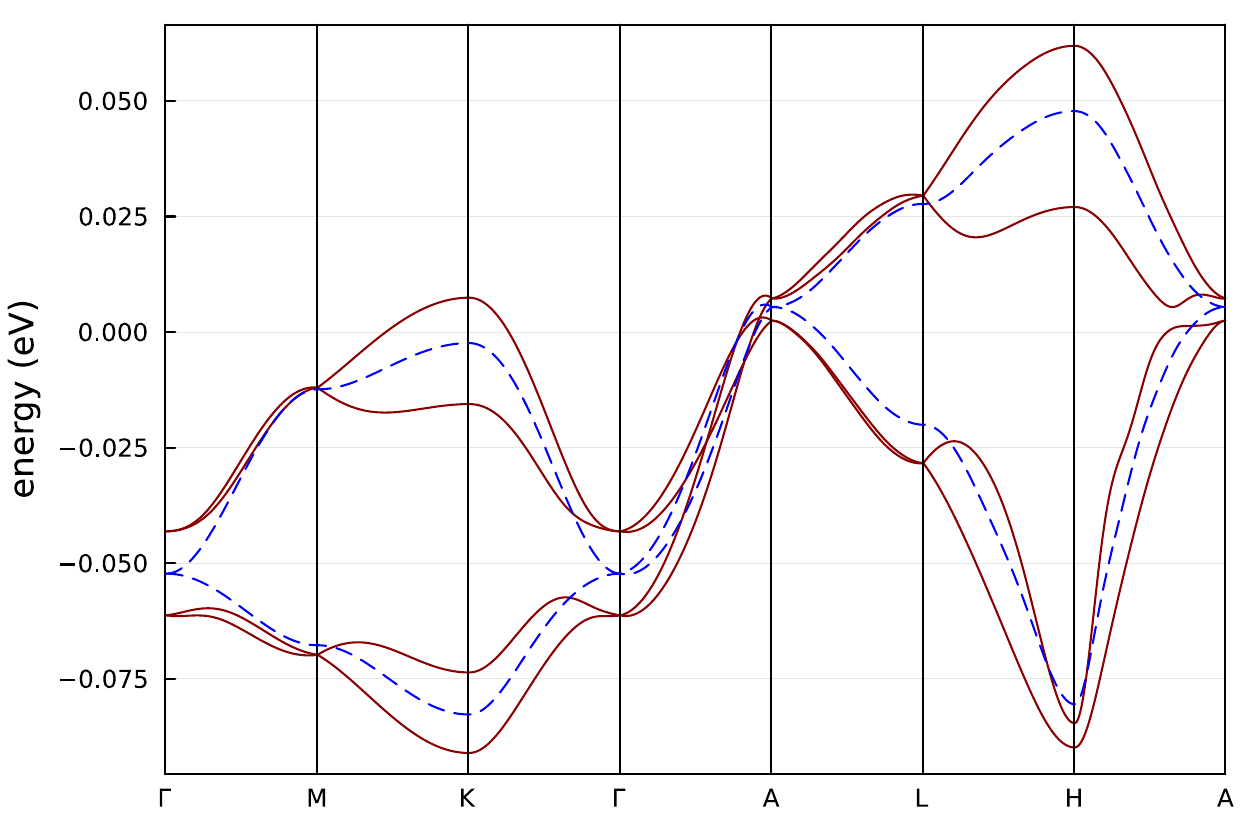}
        \caption{\label{fig:bandstructure}\label{fig:bandstructureSOC}}
    \end{subfigure}
    \caption{(a) Crystal structure of Pb$_9$Cu(PO$_4$)$_6$O in the primitive unit cell. (b) Band-structure close to the Fermi level. The blue, dashed lines depict the Cu $d_{xz}/d_{yz}$ bands obtained by a scalar-relaticvistic DFT calculation without SOC. Band crossings are enforced by symmetry at $\Gamma$ and $A$. The solid red lines are the Cu $d^{j=5/2}_{m=\pm 3/2,\pm 1/2}$ bands obtained by a full-relativistic DFT calculation. TR symmetry enforces Kramers'-degeneracy at all TRIM. Band-inversion due to SOC yields additional band crossings along $\Gamma\to A$ which are close to the Fermi surface.}
\end{figure*}

For our analysis of topological properties we start from the   relaxed crystal structure with  \emph{P3} (no.~143) space group displayed in Fig.~\ref{fig:structure}. Here, Cu and the additional O occupy positions farthest away from each other\cite{si2023}. It has been shown that the electronic structure close to the fermi energy can be effectively described by two flat bands corresponding to Cu $d_{xz}/d_{yz}$ orbitals \cite{si2023,griffin2023,lai2024,cabezas-escares2023}. 

To construct a tight-binding model that captures the topology of this two-band system, we perform a self-consistent DFT calculation and Wannier-projection onto the Cu $d_{xz}/d_{yz}$ orbitals using the %\textbf{F}ull \textbf{P}otential \textbf{L}ocal \textbf{O}rbital code \textsc{FPLO} 
full potential local orbital (\textsc{FPLO}) code
\cite{koepernik1999}. We use a dense $(11\times 11\times 15)$ $k$-mesh and the Perdew-Wang exchange-correlation potential \cite{perdew1992} for a scalar-relativistic and a fully relativistic DFT calcualtion, i.e., a DFT calculation without and with SOC. 

The obtained bandstructure is shown in figure~\ref{fig:bandstructure} and agrees well with previous results \cite{si2023,lai2024,georgescu2023,griffin2023,liu2023}. While band-crossings can be seen only at the high symmetry points $\Gamma$ and $A$ in the case without SOC (blue, dashed curve), the case with SOC (red) involves crossings at every time-reversal invariant momentum. We will discuss their topological properties in the following section. At the points $K$ and $H$ the bands are non-degenerate due to the missing inversion symmetry, and the corresponding eigenstates are $d_{xz}\pm i d_{yz}$ \cite{bauernfeind2021,eck2022}.

%\begin{figure*}
%    \centering
%    \begin{subfigure}{0.28\textwidth}
%        \includegraphics[width=\textwidth]{LK99_structure.jpg}
%        \caption{}
%    \end{subfigure}
%    \begin{subfigure}{0.35\textwidth}
%        \includegraphics[width=\textwidth]{bands_no_SOC.pdf}
%        \caption{}
%    \end{subfigure}
%    \begin{subfigure}{0.35\textwidth}
%        \includegraphics[width=\textwidth]{bands_SOC.pdf}
%        \caption{}
%    \end{subfigure}
%    %\includegraphics[width=0.45\textwidth]{bands_SOC.pdf}
%    \caption{Bands of Pb$_9$Cu(PO$_4$)$_6$O closest to the Fermi level. (a) Cu $d_{xz}/d_{yz}$ bands obtained by a scalar-relaticvistic DFT calculation. Band crossings are enforced by symmetry at $\Gamma$ and $A$.\label{fig:bandstructure} (b) Cu $d^{j=5/2}_{m=\pm 3/2,\pm 1/2}$ bands obtained by a full-relativistic DFT calculation. TR symmetry enforces Kramers'-degeneracy at all TRIM. Band-inversion due to SOC yields additional band crossings along $\Gamma\to A$ which are close to the Fermi surface.\label{fig:bandstructureSOC}}
%\end{figure*}

\section{\label{sec:Symmetry}Symmetry and Spin-Orbit coupling}
The presence of time-reversal (TR) symmetry enforces spin-degenerate bands to cross at time-reversal invariant momenta (TRIM) if these host complex valued irreducible representations of their corresponding little groups. The space-group symmetry of the crystal system is \emph{P3} (no.~143) and therefore the only point-symmetry is a $C_3$ rotation around the $z$-axis. Consequently, the little groups at any $k$-point are abelian and have one-dimensional irreducible representations. 

 The little group of the TRIM $\Gamma$ and $A$ is $C_3$ and the real valued $d_{xz}/d_{yz}$ orbitals can be decomposed into complex valued irreducible representations given by the spherical harmonics $Y_2^{\pm 1}$. Therefore, LK99 features symmetry protected band-crossings at the TRIM $\Gamma$ and $A$.

Close to a band-crossing at momentum $\mathbf{k}_0$ the Bloch-Hamiltonian can be expressed as
\begin{equation}
    \label{eq:Hamiltonian}
    H_{\mathbf{k}} =\mathbf{h}\left(\mathbf{q}\right)\cdot \boldsymbol{\sigma} + E\left(\mathbf{k}_0\right)
    \; ,
\end{equation}
where $\mathbf{h}\left(\mathbf{q}\right)$ is a smooth function of momentum $\mathbf{q}=\mathbf{k}-\mathbf{k}_0$, $\boldsymbol{\sigma}$ denotes the Pauli-matrices and $E\left(\mathbf{k}_0\right)$ is the energy at the crossing. The associated Chern number can be calculated as
\begin{equation}
    \label{eq:Chern}
    C=\frac{1}{4\pi}\int_{S^2}\!\left(\frac{\partial\hat{\mathbf{h}}\left(\mathbf{q}\right)}{\partial q^i}\times \frac{\partial\hat{\mathbf{h}}\left(\mathbf{q}\right)}{\partial q^j}\right)\cdot\hat{\mathbf{h}}\,\,\mathrm{d}q^i\!\!\wedge\! \mathrm{d}q^j \; .
\end{equation}
Here and in the following, sum convention is implied, $\hat{\mathbf{h}}\left(\mathbf{q}\right) = \frac{\mathbf{h}\left(\mathbf{q}\right)}{\lVert \mathbf{h}\left(\mathbf{q}\right)\rVert}$ and $S^2$ denotes a sphere surrounding the Weyl point. In the language of differential topology, this Chern number is just the degree of the smooth map $\hat{\mathbf{h}}:S^2\to S^2$ which we can determine via
\begin{equation}
    \label{eq:degree}
    C\equiv\mathrm{deg}\left(\hat{\mathbf{h}}\right)=\sum_{\mathbf{q}\in\hat{\mathbf{h}}^{-1}(\boldsymbol{\eta})}\mathrm{sign}\left[\mathrm{det}\left(\mathrm{d}\hat{\mathbf{h}}\vert_\mathbf{q}\right)\right]
\end{equation}
where $\mathrm{d}\hat{\mathbf{h}}\vert_\mathbf{q}$ is the differential (i.e. Jacobian) of $\hat{\mathbf{h}}$ evaluated at a momentum $\mathbf{q}$ from the pre-image $\hat{\mathbf{h}}^{-1}(\boldsymbol{\eta})$ of any regular value $\boldsymbol{\eta}$ of $\hat{\mathbf{h}}$. 

A band crossing at $\mathbf{k}_0$ is a Weyl point if and only if the Chern number and so the degree of $\hhat$ is non-zero. When moving away from the crossing, the bands generically have to split in every direction due to the low symmetry of our system. This favors the crossings to be Weyl-points and indeed a numerical calculation confirms this. For this low symmetry, a Chern number being zero would be accidental in the sense that one could always introduce a small perturbation allowed by symmetry such that one would get a non-zero Chern number. Therefore, and because we checked numerically, we will assume in the following that the considered band-crossings have non-vanishing Chern number and are hence Weyl points.
%\khc{Did you check the Chern number afterwards, then let's say so, or does the numerical approach with the vector field warranty it is is Weyl point, think it does, then let's say so here. With SOC and low symmetry it might indeed be that crossing points are usually Weyl points, but not in general. At least at some higher symmetry momenta, I could imagine that the spin of the two crossing bands point in the same direction and then you don't have a Weyl point.}

Time-reversal and $C_3$ symmetry almost completely determine the degree of $\hhat$ up to its sign as we explain in appendix \ref{app:Symmetry}. There we find that, for the TRIM $\Gamma$ and $A$ in the case without SOC, the map $\hat{\mathbf{h}}$ has degree two, since TR enforces that $\hat{h}_{x/y}(\mathbf{q})=\hat{h}_{x/y}(\mathbf{-q})$. Consequently, $\hat{\mathbf{h}}$ wraps around the sphere twice and Chern numbers at $\Gamma$ and $A$ have absolute values equal to two \cite{hirschmann2023,tsirkin2017}. To determine the sign, we calculate Eq.~(\ref{eq:Chern}) numerically and obtain Chern numbers equal to $-2$ at $\Gamma$ and $+2$ at $A$ respectively.

The situation changes if we take SOC into account. Then each of the two bands splits, yielding four bands in total as shown in Fig.~\ref{fig:bandstructureSOC}, where we present the band-structure of a full-relativistic DFT calculation (red curve). These bands belong to those copper $3d_{5/2}$ orbitals which have $j_z=\pm1/2$ and $j_z=\pm3/2$. However, now having a spin-full representation of the corresponding double group, the TR operator squares to $-1$ thereby enforcing Kramers' degeneracies of the bands at every TRIM. Hence, there are further Weyl points at $M$ and $L$ additional to those at $\Gamma$ and $A$. Furthermore, this is valid for both pairs of bands, which implies that every TRIM holds two Weyl points at different energies. This is especially interesting for $A$ where Weyl points are very close to the Fermi energy $\epsilon_F=0$ as can be seen in Fig.~\ref{fig:bandstructureSOC}.

%\begin{figure}
 %   \centering
  %  \includegraphics[width=\columnwidth]{bands_SOC.pdf}
  %  \caption{Cu $d^{j=5/2}_{m=\pm 3/2,\pm 1/2}$ bands obtained by a full-relativistic DFT calculation.}
   % \label{fig:bandstructureSOC}
%\end{figure}

At $M$ and $L$ the Chern numbers must have an absolute value equal to one, since the corresponding little groups are the trivial group. At $\Gamma$ and $A$ the Chern numbers depend on the character of the bands. Due to the presence of time-reversal symmetry the degenerate bands have either $j_z=\pm \frac{1}{2}$ or $j_z=\pm \frac{3}{2}$. In the former case $\hat{\mathbf{h}}$ transforms like a 3D vector under $C_3$, and hence the absolute value of the corresponding Chern number is one. In the latter case, $\hhat$ is invariant under $C_3$. Hence, for any regular value of $\hhat$ the pre-image contains three points at which the change of orientation is the same and hence, Eq.~(\ref{eq:degree}) implies Chern numbers equal to plus or minus three. Again, the details of the symmetry analysis for the case with SOC can be found in appendix \ref{app:Symmetry}.

However, only a direct calculation can determine the characters of the bands. From our full relativistic DFT calculation we find that at $\Gamma$ the upper band is of $\pm\frac{3}{2}$ character and the lower one $\pm\frac{1}{2}$. Hence we obtain Chern numbers equal to $C=+3$ and $C=-1$ respectively. At $A$ the situation is reversed yielding $C=+1$ for the upper and $C=-3$ for the lower band.

At $M$ we get a Chern number $C=-1$ and $C=+1$ at $L$ for the upper bands. For the lower ones they are $C=+1$ at both $M$ and $L$. Hence, if we sum up the Chern numbers of the upper band we arrive at a total of $+4$ and $+2$ for the lower band (note that $L$ and $M$ are three-fold, due to the $C_3$ axis). Since the Nielsen–Ninomiya theorem enforces a total of zero Chern numbers \cite{nielsen1981,nielsen1981a}, there must be additional Weyl points, not protected by symmetry.

In figure \ref{fig:bandstructureSOC} we can see band-crossings at the TRIM, but also at the line from $\Gamma$ to $A$. The latter crossings are close to the Fermi energy $\epsilon_F=0$ and also somewhat below. Due to TR invariance the same crossings must appear at the corresponding negative momenta on the line from $\Gamma$ to $-A$. We calculate the Chern numbers of these crossing-points. For the lowest band there is one Weyl point with $C=-1$ and its time-reversal partner with the same Chern number. Thus, for the lowest band the total of Chern numbers vanishes. For the remaining bands, we do the same calculation, but find that the total does not equal zero. Consequently there must be additional Weyl points at general momenta that are not on the 
high symmetry k-path shown in figure \ref{fig:bandstructureSOC}. We have to find them manually.

\section{\label{sec:Algorithm}An algorithm to detect Weyl points}
Finding Weyl points whose positions are not determined by symmetry arguments can be a peculiar task. Typical approaches divide the Brillouin-zone into multiple parallelepipeds, integrate the Berry-curvature over their surfaces, and refine the enclosed volume if it has a non-vanishing Chern number \cite{xu2020}. Alternatively, there is a method based on Wilson-loops \cite{saini2022a} and on a direct search for local minima in the band-gap \cite{wu2018}. Here, we propose a completely different approach that can be used complementary to the others. It is based on the fact that Weyl points act as sources and sinks of Berry-curvature:
\begin{equation}
\label{eq:berrycurvature}
    \Omega^n_{ij}=-\mathrm{Im}\left[\sum_{m\neq n}\frac{\braket{\psi^n_k\lvert \partial_{k^i}H(k) \rvert\psi^m_k}\braket{\psi^m_k\lvert \partial_{k^j}H(k) \rvert\psi^n_k}}{\left(E_m(k)-E_n(k)\right)^2}\right].
\end{equation}
Here, $\psi^n_k$ denotes an eigenstate of the Bloch-Hamiltonian $H(k)$ with band-index $n$, momentum $k$ and energy $E_n(k)$.

More precisely, for a fixed band-index $n$ we can understand the Berry-curvature (Eq.~\ref{eq:berrycurvature}) as a differential two-form $\Omega^n:=\Omega^n_{ij}\mathrm{d}k^i\wedge\mathrm{d}k^j$ on the Brillouin-zone without the Weyl points. In 3D we can use the Hodge-$\star$ to identify $\Omega^n$ with the one-form $\star\Omega^n=\Omega^n_{ij}\epsilon_{ijl}\mathrm{d}k^l=:X_l \mathrm{d}k^l$ and use the canonical isomorphism of tangent and co-tangent space to finally identify it with a vector-field $X=X^i\partial_{k^i}$. 

To find a Weyl point, we pick a starting-point $k$ in the Brillouin-zone and search for an integral curve $\gamma(t)$ of $X$ with $\gamma(0)=k$. That is, we solve the ordinary differential equation
\begin{equation}
\label{eq:WeylODE}
    \frac{\mathrm{d}\gamma}{\mathrm{d}t} = X(\gamma(t)).
\end{equation}
If the solution converges to a point in finite time, we have found a band-crossing, which acts as a sink of the vector field $X$. Otherwise, we may get a closed curve, and then disregard it and pick another starting point. 

To find all points we apply this procedure to multiple starting vectors distributed over the whole Brillouin-zone. This is trivially parallelizable. In order to also find sources instead of sinks, we apply the procedure to $-X$. For the numerical solution of Eq. \ref{eq:WeylODE} we apply the implicit Euler-method with adaptive step size \cite{rackauckas2017}. %This way we are able to detect 32 Weyl points as described in the next section.

\begin{figure}
    \centering
    \includegraphics[width=\columnwidth]{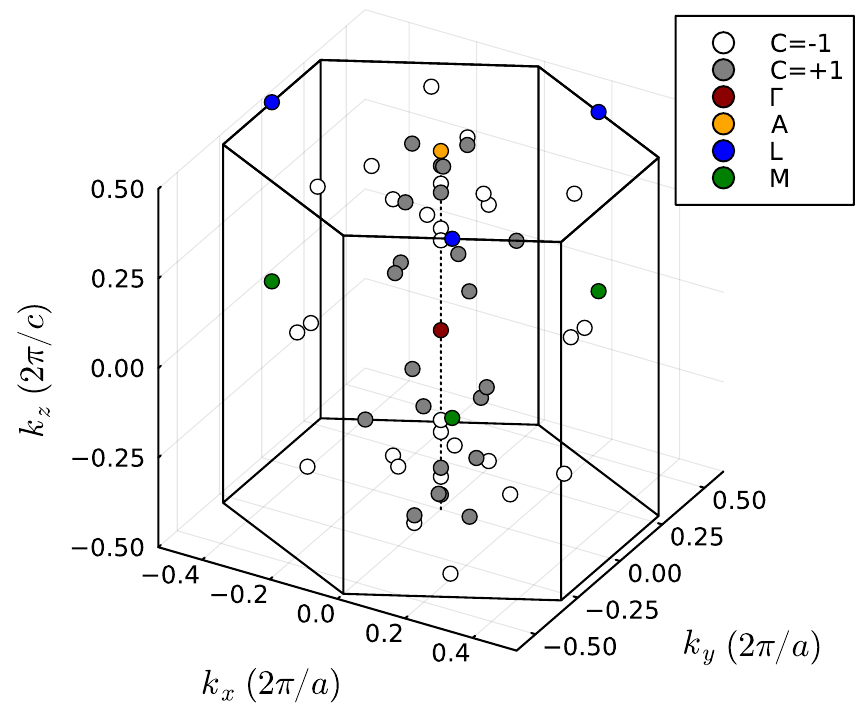}
    \caption{Weyl points in the Brillouin-zone. The color-code denotes the different Chern numbers $C$. At TRIM, the Chern numbers are different for the $j_z=\pm\frac{1}{2}$ and $j_z=\pm\frac{3}{2}$ bands (see text).}
    \label{fig:weylpoints}
\end{figure}

Thus, we can find the remaining Weyl points. For the upper pair of bands, we detect 32 and 34 for the lower pair of bands (including those described earlier) which are shown in Fig.~\ref{fig:weylpoints}. To confirm the validity of the algorithm, we calculate the Chern numbers of the points via an integration of Eq.~(\ref{eq:berrycurvature}) over a small sphere surrounding a point. This confirmed that we did not get false positives. We also sum up all Chern numbers yielding zero as it must be. Our algorithm detected all Weyl points at TRIM and also those at general momenta without any guidance. Hence, it is suitable for the automatic detection of Weyl points.

\section{\label{sec:Surface}Surface states}

\begin{figure*}
    \centering
    \begin{subfigure}{0.49\textwidth}
        \includegraphics[width=\textwidth]{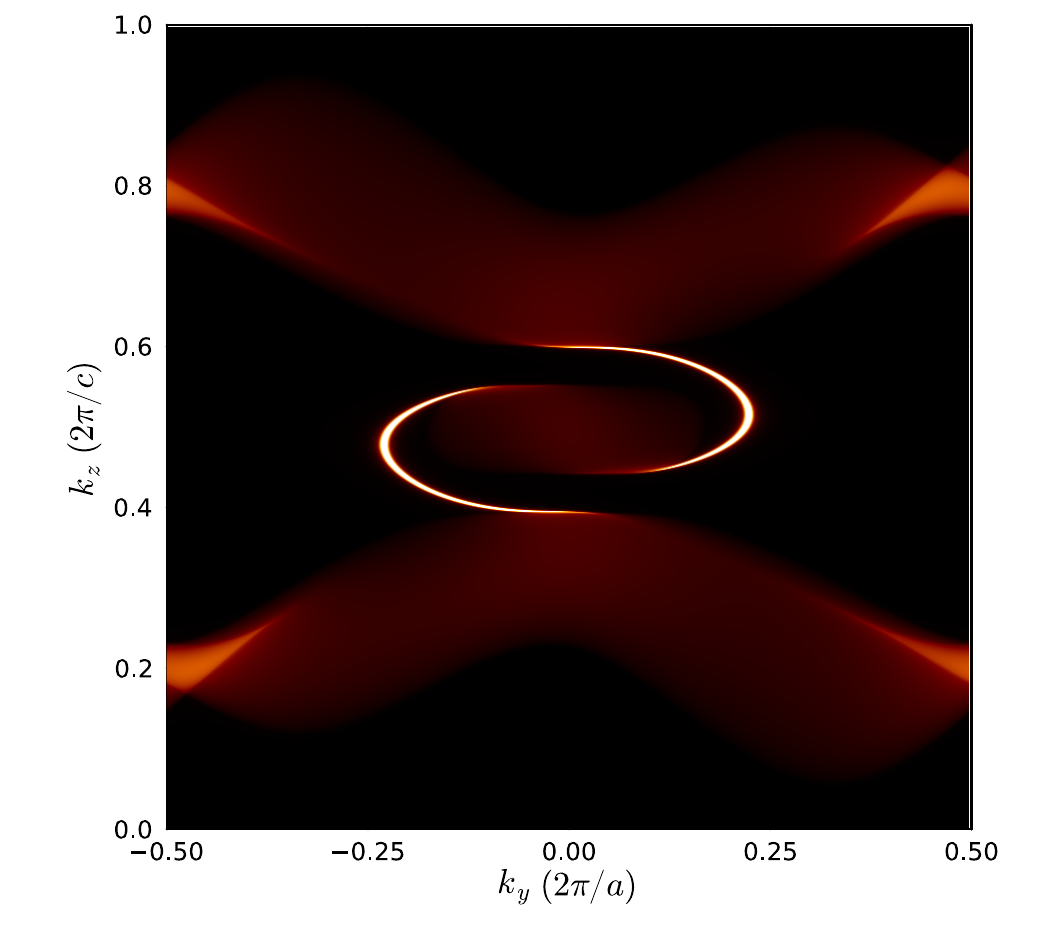}
        \caption{\label{fig:surf_spec_dens}}
    \end{subfigure}
    \begin{subfigure}{0.49\textwidth}
        \includegraphics[width=\textwidth]{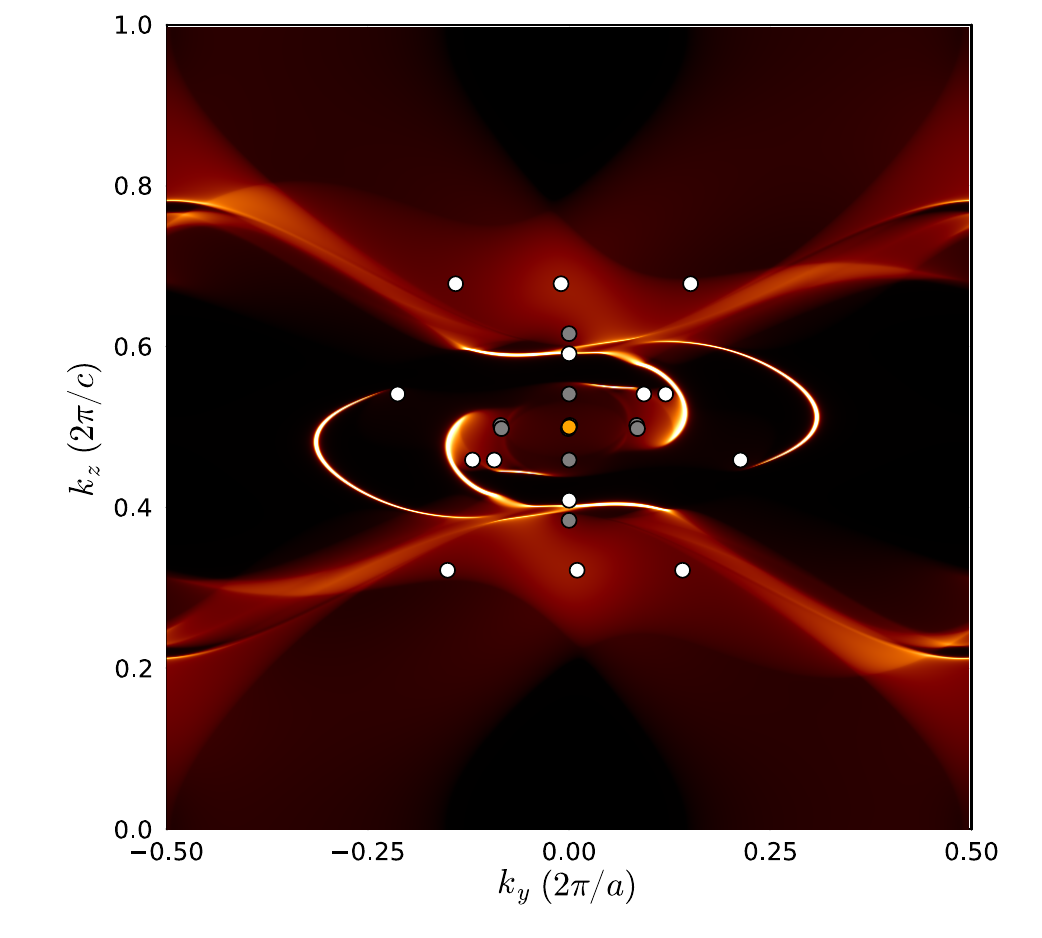}
        \caption{\label{fig:surf_spec_dens_SOC}}
    \end{subfigure}
    \caption{Surface spectral density $A(k_y,k_z,\omega=\epsilon_F)$ calculated from Eq.~(\ref{eq:surf_spec_dens}) for a semi-infinite crystal terminated at the $x=0$ surface. (a) Without SOC there are two Fermi-arcs connecting those parts of the Fermi surface that surround the projections of Weyl points at $\Gamma$ and $A$ with Chern numbers equal to -2 and +2 respectively. (b) SOC leads to the appearance of further Fermi-arcs. The Weyl points are color-coded as in Fig.~\ref{fig:weylpoints}, but we show only those close to the Fermi level. Two Weyl points with $C=+1$ are projected onto the same point as $A$ (orange). To the left and right of $A$, Weyl points (grey) with $C=+1$ are projected onto the same point as well.}
    
\end{figure*}

Due to the bulk-boundary-correspondence, the presence of Weyl points implies the existence of Fermi-arcs, i.e., states of constant energy that are localized on a surface of the crystal. They appear if the corresponding reciprocal surface in $k$-space contains  projections of Weyl points with opposite chiralities that do not fall on the same point\cite{wan2011}. Fermi-arcs can be determined by angle resolved photo emission spectroscopy (ARPES)\cite{belopolski2016,armitage2018} and can be calculated via the surface spectral density of a semi-infinite crystal
\begin{equation}
\label{eq:surf_spec_dens}
    A(k_y,k_z,\omega) = -\mathrm{Im}\left[\mathrm{Tr}\left(\frac{1}{\omega+i0^+-H(k_y,k_z)}\right)\right].
\end{equation}
Here we choose to terminate the crystal at the $x=0$ surface. Hence, $H(k_y,k_z)$ denotes the Fourier-transformed tight-binding Hamiltonian where hopping is retained in the positive $x$-direction only. The trace Tr runs over all states localized at the surface. For numerical calculations we truncate the system after 1344 layers and choose the positive infinitesimal to be $0^+=10^{-4}$.

Fig.~\ref{fig:surf_spec_dens} shows the projected Fermi surface and Fermi-arcs for our calculation without SOC. The Fermi surface consists of three disconnected parts. The central one encloses the projection of the Weyl point at $A$. Due to periodicity of the Brillouin-zone, the other two parts sandwich the projection of the Weyl point at $\Gamma$. Since the Chern number of the  Weyl point at A in the center ($k_z=1/2$) is $+2$ and that of $\Gamma$ at the top or bottom ($k_z=0,1$) is $-2$, there are two arcs connecting the different parts of the Fermi surface. Note, the arcs do not exactly terminate at A and $\Gamma$ because these momenta are not precisely at the Fermi energy.

If we consider the full-relativistic calculation, that includes SOC, the situation becomes more complex due to the presence of multiple Weyl points as shown in Fig.~\ref{fig:surf_spec_dens_SOC}. Due to SOC the bands are wider and the projected Fermi surface merges in $k_z$-direction, such that compared to the case without SOC we only have two disconnected parts: a central one and the merged top/bottom part. Outside those parts we observe two isolated Weyl points (white circles) which are essentially at the Fermi level. Therefore, from each of them one arc emerges and connects the Weyl points to the top/bottom part of the Fermi surface. 

The central part encloses multiple projections of Weyl points which are a bit farther away from the Fermi energy. Hence, we observe two additional arcs that do not start at a Weyl point, but connect the central part with the top/bottom part of the Fermi surface. Tangential\cite{haldane2014} at the central part we can see two further arcs, which are small and almost horizontal. They connect Weyl points of opposite chirality (white and grey), that are both projected onto the same component of the Fermi surface. For all these arcs we confirm their topological nature by calculating the surface-projected bandstructure along loops surrounding a Weyl point and observe topological edge states that connect the upper bulk-projected band with the lower one. Hence, compared to Fig.~\ref{fig:surf_spec_dens}, SOC enriches the Fermi surface with additional topologically protected states.

\begin{figure*}
    \centering
    \begin{subfigure}{0.48\textwidth}
    \caption{$\epsilon_F=-57$ meV\label{fig:surf_spec_spin_dens_a}}
    \includegraphics[width=\textwidth]{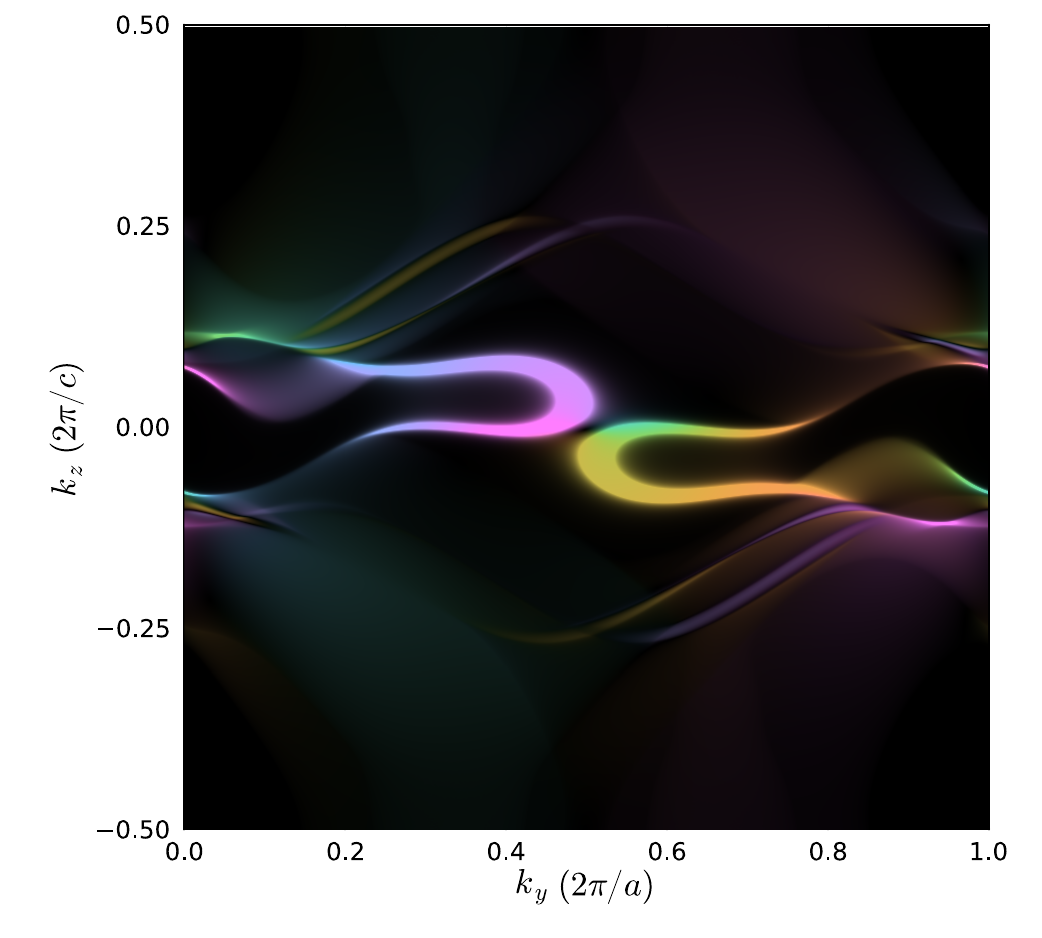}
    \end{subfigure}
    \begin{subfigure}{0.48\textwidth}
    \caption{$\epsilon_F=-56.5$ meV\label{fig:surf_spec_spin_dens_b}}
    \includegraphics[width=\textwidth]{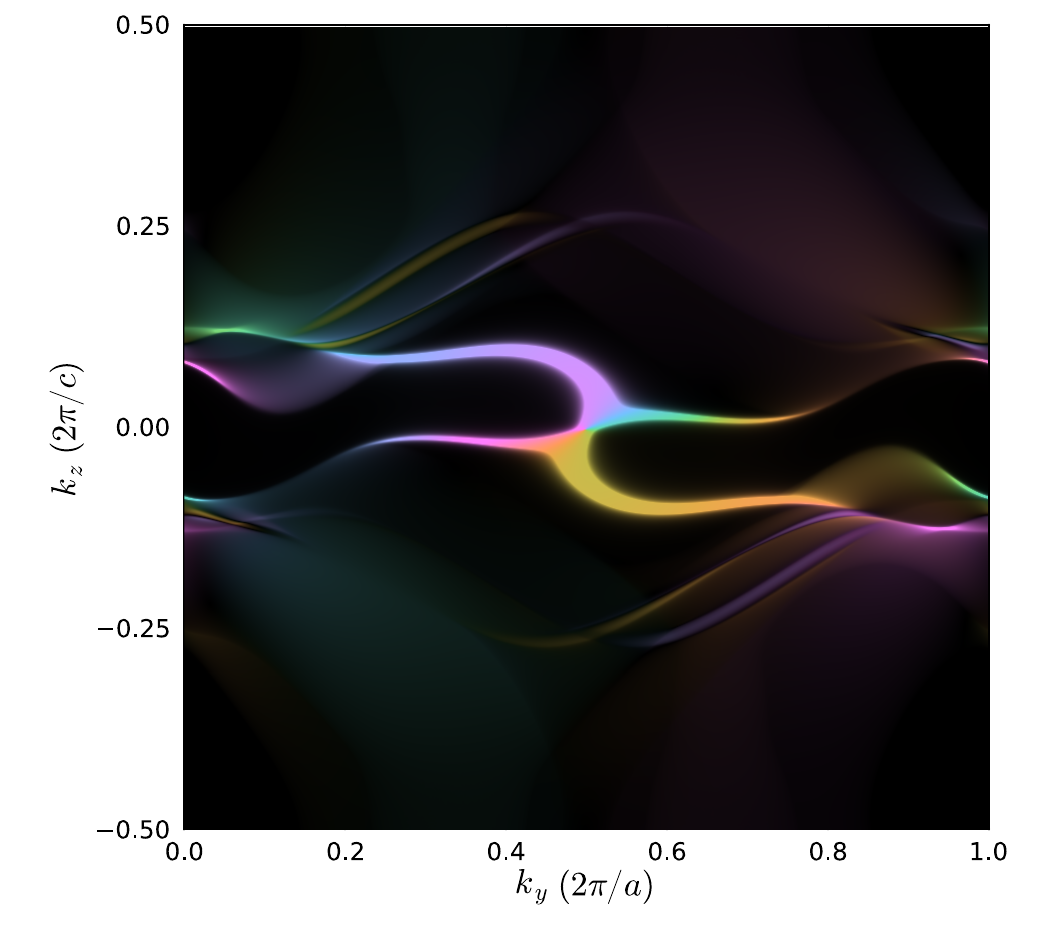}
    \end{subfigure}
    \begin{subfigure}{0.48\textwidth}
    \caption{$\epsilon_F=-55$ meV\label{fig:surf_spec_spin_dens_c}}
    \includegraphics[width=\textwidth]{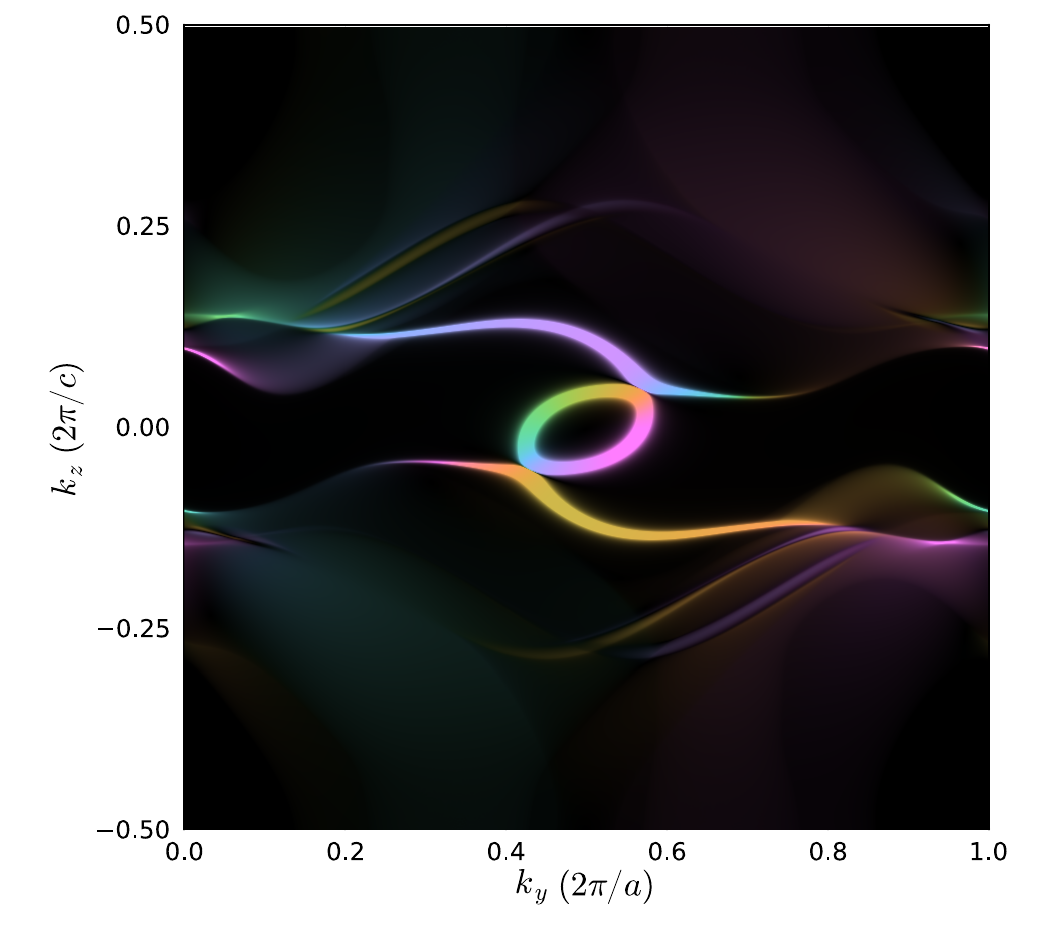}
    \end{subfigure}
    \begin{subfigure}{0.48\textwidth}
    \caption{$\epsilon_F=-52$ meV\label{fig:surf_spec_spin_dens_d}}
    \includegraphics[width=\textwidth]{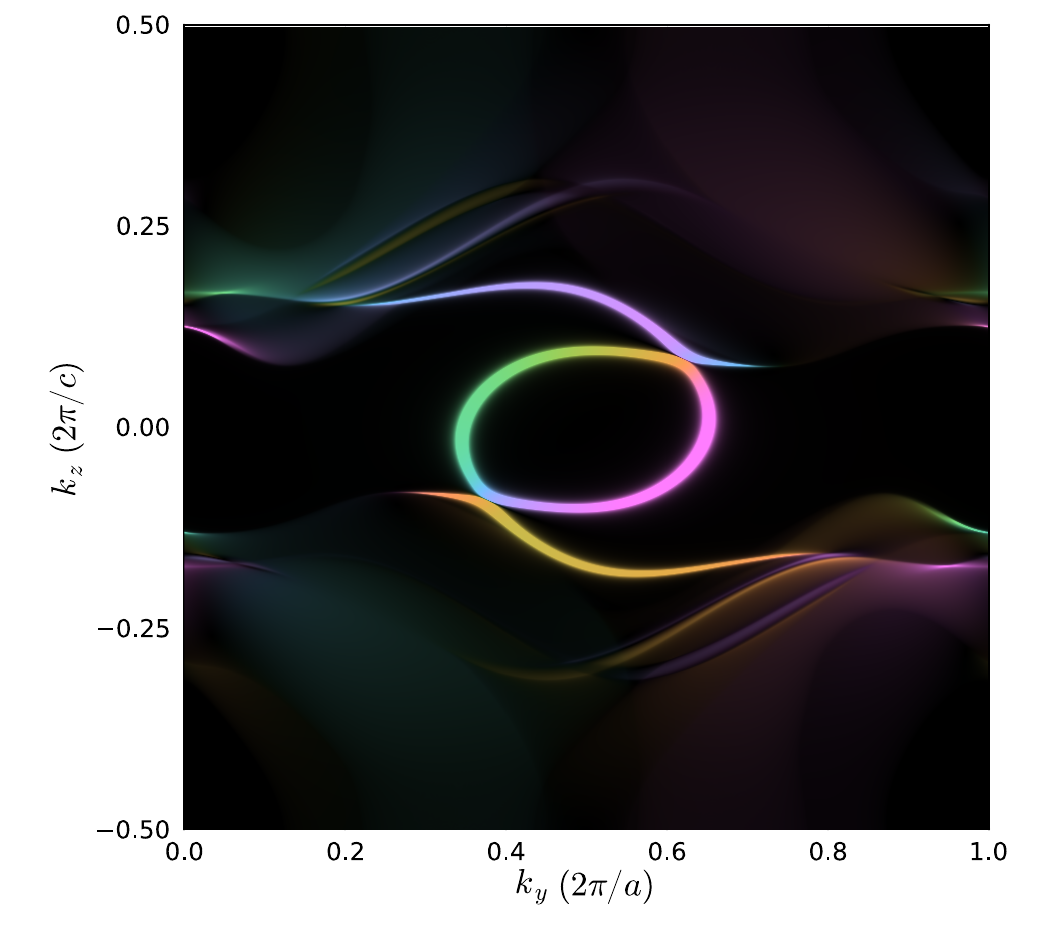}
    \end{subfigure}
    \includegraphics[width=0.3\textwidth]{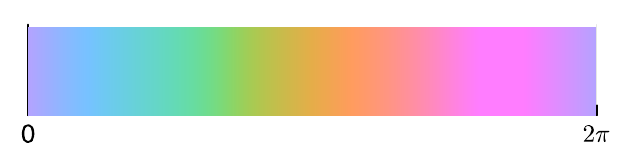}
    \caption{Surface spectral spin-density $\mathbf{P}(k_y,k_z,\omega=\epsilon_F)$ caclualted from Eq.~(\ref{eq:surf_spec_spin_dens}) for a semi-infinite crystal terminated at the $x=0$ surface. Colors represent the angles of spin in the yz-plane (see color bar at the bottom), brightness corresponds to magnitude $\lVert \mathbf{P}\rVert$. When increasing the Fermi energy from (a) to (b) the two Fermi-arcs touch and the spin-polarization winds around the touching point. A further increase leads to avoided crossings (c) and a surface-loop which is detached from the Weyl points. The spin-polarization winds around the loop which becomes larger for higher $\epsilon_F$ (d).}
    \label{fig:surf_spec_spin_dens}
\end{figure*}

Furthermore, SOC can lead to interesting patterns in the spin-polarization, or spin-texture of a surface. This can be measured with spin-resolved ARPES\cite{hoesch2002,lv2019}. In the following, we study the spin-polarization of the surface states at different Fermi levels. For the same semi-infinite crystal as above, we can calculate the surface spin-polarization via
\begin{equation}
\label{eq:surf_spec_spin_dens}
    \mathbf{P}(k_y,k_z,\omega) = -\mathrm{Im}\left[\mathrm{Tr}\left(\frac{\mathbf{S}}{\omega+i0^+-H(k_y,k_z)}\right)\right]
\end{equation}
where $\mathbf{S}$ denotes the spin operator and the trace is again taken over all states localized at the surface. The results are shown in Fig. \ref{fig:surf_spec_spin_dens}.

At a Fermi energy equal to $-57.0$ meV in Fig.~\ref{fig:surf_spec_spin_dens_a}, we see two U-shaped surface arcs, where the spin-projection onto the yz-plane points into different directions for the arcs. If the Fermi energy is increased to $-56.5$ meV (\subref{fig:surf_spec_spin_dens_b}) the arcs touch and the spin-projection winds around the touching point. Upon further increase in the Fermi energy (\subref{fig:surf_spec_spin_dens_c}, \subref{fig:surf_spec_spin_dens_d}) we observe avoided crossings which turn the two U-shaped arcs into two horizontal arcs and a detached loop in the center. The spin-projection winds around this loop and near the avoided crossings we can observe changes in polarization as expected in such a situation. %The loop surrounds the projection of the two M-points ($(1/2,1/2,0)$ and $(0,1/2,0)$ in the primitive basis) with a total Cern-number equal to $+2$. \mb{for the lower bands, $-2$ for the upper}

An early study proposed that such a detached loop reflects the topological invariant associated with a transition between normal and topological insulators via a Weyl-semi-metal phase\cite{huang2015}. A different publication\cite{devizorova2017} attributed the emergence of a loop to a delicate interplay between inter- and intra-valley interactions in a four-valley model for Weyl-semi-metals. Furthermore, contractible loops on the Fermi surface have also been interpreted as track-states\cite{mccormick2017} or as gaped surface states as a result of quadrupole topology in higher order Weyl-semi-metals\cite{wang2020}. Here we demonstrated the presence of a detached loop in an ab-initio derived tight-binding model. To verify whether it is related to the latter concept, further calculations are necessary.

\section{\label{sec:Conclusions}Conclusion}
We have shown that SOC significantly modifies the band-structure of Pb$_9$Cu(PO$_4$)$_6$O leading to the presence of 66 Weyl points and developed an algorithm to efficiently identify their positions. Furthermore, we described the corresponding surface states and found that depending on the actual value of the Fermi energy these can show different features. Especially an avoided crossing is possible that influences the spin-polarization of surface states and results in a detached loop not connecting any Weyl points.

The very complex and beautiful Weyl physics described here should be observable in Pb$_9$Cu(PO$_4$)$_6$O$_{1-\delta}$ if the material is cooled to a few Kelvin so that the Cu sites order in a long-range pattern. Further, an oxygen off-stoichiometry ($\delta\neq0$) or another means of electron or hole doping of the  Mott insulator is needed to turn it into a metal. Different $\delta$'s then correspond to different Fermi energies in our theoretical calculation. All bands will be (quasiparticle-)renormalized, i.e., reduced in width, compared to the DFT description of the present paper.

\begin{acknowledgments}
We would like to thank Markus Wallerberger and Giorgio Sangiovanni for many fruitful discussions. Further we acknowledge funding through the Austrian Science Fund (FWF) projects I 5398, P 36213, P 33571, SFB Q-M\&S (FWF project ID F86), and Research Unit QUAST by the Deutsche Foschungsgemeinschaft (DFG; project ID FOR5249) and FWF (project ID I 5868). L.~S.~is thankful for the starting funds from Northwest University. Calculations have been done in part on the Vienna Scientific Cluster (VSC). 

For the purpose of open access, the authors have applied a CC BY public copyright licence to any Author Accepted Manuscript version arising from this submission.

The data of our calculations and the code for the detection of Weyl points are openly available  at XXX and XXX, respectively.

\end{acknowledgments}

\appendix
\section{Symmetry\label{app:Symmetry}}
For a Weyl point at a TRIM the absolute value of its Chern number can be derived from symmetry arguments as we have done in Section \ref{sec:Symmetry}. Here we provide the details of these arguments starting with the case without SOC, where the TR operator can be represented as $\Theta:=e^{-i\pi L_y}\mathcal{K}$. $L_y$ is the $y$-component of orbital angular momentum and $\mathcal{K}$ denotes complex conjugation. In the basis of the spherical harmonics $Y_2^{\pm1}$ it can thus be expressed as
\begin{equation}
    \label{eqTRnoS}
    \Theta = \left(\begin{array}{cc}
         0 & -1 \\
         -1 & 0
    \end{array}\right) \mathcal{K}.
\end{equation}
This allows us to directly calculate how the Pauli matrices in Eq.~(\ref{eq:Hamiltonian}) transform under TR, if we choose the basis of $Y_2^{\pm1}$ for the $d_{xz}/d_{yz}$ orbitals, which is much more practical for the following calculations\kh{:} $\boldsymbol{\sigma}_x$ and $\boldsymbol{\sigma}_y$ stay invariant and $\boldsymbol{\sigma}_z$ changes sign. This implies that the $x$ and $y$ components of $\hat{\mathbf{h}}(\mathbf{q})$ are even functions of $\mathbf{q}$ and $\hat{h}_z(\mathbf{q})$ is odd, i.e.
\begin{equation}
\label{eq:TR_h_no_SOC}
    \hat{\mathbf{h}}(\mathbf{q}) := \left(\begin{array}{c}
        \hat{h}_x(\mathbf{q})\\ \hat{h}_y(\mathbf{q})\\ \hat{h}_z(\mathbf{q})
    \end{array}\right)= \left(\begin{array}{c}
        \hat{h}_x(-\mathbf{q})\\ \hat{h}_y(-\mathbf{q})\\ -\hat{h}_z(\mathbf{-q})
    \end{array}\right).
\end{equation}

The degree of $\hat{\mathbf{h}}:S^2\to S^2$ as a smooth map from a sphere onto a sphere can be determined with the help of Eq.~(\ref{eq:degree}) at any regular value $\boldsymbol{\eta}$ of $\hhat$. $\boldsymbol{\eta}$ is called a regular value of $\hat{\mathbf{h}}$ if its pre-image $\hat{\mathbf{h}}^{-1}(\boldsymbol{\eta})=\lbrace \mathbf{q}_1,\hdots ,\mathbf{q}_r\rbrace$ consists of points at which the differential of $\hat{\mathbf{h}}$ has full rank. If $s$ denotes the number of points in $\hat{\mathbf{h}}^{-1}(\boldsymbol{\eta})$ at which $\hat{\mathbf{h}}$ changes orientation, then the degree of $\hat{\mathbf{h}}$ equals $r-s$ [this is just a reformulation of Eq.~(\ref{eq:degree})]. As we have stated earlier, we are considering band-crossings with $C\neq0$ only, i.e. Weyl points. Then every value of $\hhat$ is regular.

If we pick a regular value $\boldsymbol{\eta}=\hhat(\mathbf{q}_0)$ at which $\hat{h}_z(\mathbf{q}_0) = 0$, we can conclude from Eq.~(\ref{eq:TR_h_no_SOC}) that the pre-image $\hhat^{-1}(\boldsymbol{\eta})$ contains at least two points $\lbrace \mathbf{q}_0,-\mathbf{q}_0\rbrace$ (at which the change in orientation is the same as $\hhat:S^2\to S^2$ smoothly maps a sphere onto a sphere). Hence, the degree of $\hhat$ is at least $\pm 2$. In principle every even number would be allowed by symmetry, but one could always introduce a small perturbation that is consistent with symmetry such that the degree would become $\pm2$. Thus, we conclude that the Weyl point's Chern number is $\pm 2$ without SOC. The sign cannot be determined by symmetry only.

%Using the calculus of characters we can determine which of the $Y_l^m$ form an $A$ irrep and which form the $E$ representation. We find that for $l=1$ the $A$ irrep is spanned by $Y_1^0$ and for $l=2$ the $E$ representation is spanned by $Y_2^{\pm 2}(\mathbf{q})\sim e^{\pm i2q_\phi}$ where $q_\phi$ is the azimuthal angle. 

%For our Weyl points at $\Gamma$ and $A$ we look at a momentum with $\lVert\mathbf{q}\rVert>0$, $q_\phi = 0$ and $q_\theta=\pi/2$. As the differential of $\hat{\mathbf{h}}(\mathbf{q})$ has full rank at this point it is a regular point. As the $x$ and $y$ components contain terms with even $l$ and the lowest order ones have $\lvert m\rvert=2$, we conclude that $\hat{\mathbf{h}}(q_\theta=\frac{\pi}{2},q_\phi=0)=\hat{\mathbf{h}}(q_\theta=\frac{\pi}{2},q_\phi=\pi)$. A direct computation reveals, that also the second momentum with $(q_\theta=\frac{\pi}{2},q_\phi=\pi)$ is regular. Hence, $\hat{\mathbf{h}}(q_\theta=\frac{\pi}{2},q_\phi=0)$ is a regular value of $\hat{\mathbf{h}}$ with two points in its pre-image. Whether $\hat{\mathbf{h}}$ changes orientation on these points depends on the signs of the expansion coefficients, however orientation will be the same on both points. Hence, $\hat{\mathbf{h}}(\mathbf{q})$ wraps around the sphere twice and its degree has absolute value equal to two.

For spin-full bands, i.e., with SOC, the situation is different, as the TR operator can be expressed as $\Theta = e^{-i\pi J_y}\mathcal{K}$ with total angular momentum $J_y$ which has half-integer eigenvalues as opposed to $L_y$. On the basis of the $(j=\frac{5}{2},j_z=\pm\frac{1}{2})$ orbitals this takes the form
\begin{equation}
    \label{eq:TR_S}
    \Theta = \left(\begin{array}{cc}
        0 & -1 \\
        1 &0 
    \end{array}\right)\mathcal{K} \; ;
\end{equation}
and up to an overall minus sign this is the same for the $(j=\frac{5}{2},j_z=\pm\frac{3}{2})$ orbitals. Again, we can use this to determine how the Pauli matrices transform under TR. Now all of them change sign, which implies that every component of $\hat{\mathbf{h}}(\mathbf{q})$ is an odd function of momentum $\mathbf{q}$. 

If there is only the trivial point-group present at a Weyl point as it is the case for the TRIM $L$ and $M$, then the most generic Chern number is $\pm 1$. However, at $\Gamma$ and $A$ the point group symmetry can give additional constraints on $\hhat$ depending on the representation of the orbitals.

The $(j=\frac{5}{2},j_z=\pm\frac{1}{2})$ orbitals form a representation of the double cover of $C_3$ which implies that $\hhat(\mathbf{q})$ transforms like a 3D vector under the $C_3$ rotation. Therefore, the most generic $\hhat$ consistent with TR and $C_3$ symmetry wraps around the sphere only once and hence a Weyl point with this orbital character would have a Chern number equal to $\pm 1$.

The $(j=\frac{5}{2},j_z=\pm\frac{3}{2})$ orbitals form a representation of the double cover of $C_3$ which leads to different point group transformation properties of $\mathbf{h}(\mathbf{q})$, namely all three components transform trivially, i.e., they do not change at all. This implies for any regular value $\boldsymbol{\eta}$ there are at least three points in its pre-image (and again the change of orientation is the same for each of them). Therefore, the most generic $\hhat$ consistent with TR and $C_3$ symmetry wraps around the sphere three times and hence a Weyl point with this orbital character would have a Chern number equal to $\pm 3$.

%To determine the degree of $\mathbf{h}(\mathbf{q})$, it is not sufficient to look at the $l=1$ terms, as there is only $Y_1^0$ that contributes, but does not depend on $q_\phi$ (i.e. considering only $Y_1^0$ the map $\hat{\mathbf{h}}$ would have no regular points). However, with the help of characters one can confirm that a linear combination of $Y_3^{\pm 3}\sim e^{\pm 3iq_\phi}$ forms an $A$ irrep. Therefore, the degree of $\mathbf{h}(\mathbf{q})$ has an absolute value equal to three.

%\AtEveryBibitem{\clearfield{title}}
%\bibliographystyle{apsrev4-2}
\bibliography{Weyl-points_in_LK99}

\begin{thebibliography}{43}%
\makeatletter
\providecommand \@ifxundefined [1]{%
 \@ifx{#1\undefined}
}%
\providecommand \@ifnum [1]{%
 \ifnum #1\expandafter \@firstoftwo
 \else \expandafter \@secondoftwo
 \fi
}%
\providecommand \@ifx [1]{%
 \ifx #1\expandafter \@firstoftwo
 \else \expandafter \@secondoftwo
 \fi
}%
\providecommand \natexlab [1]{#1}%
\providecommand \enquote  [1]{``#1''}%
\providecommand \bibnamefont  [1]{#1}%
\providecommand \bibfnamefont [1]{#1}%
\providecommand \citenamefont [1]{#1}%
\providecommand \href@noop [0]{\@secondoftwo}%
\providecommand \href [0]{\begingroup \@sanitize@url \@href}%
\providecommand \@href[1]{\@@startlink{#1}\@@href}%
\providecommand \@@href[1]{\endgroup#1\@@endlink}%
\providecommand \@sanitize@url [0]{\catcode `\\12\catcode `\$12\catcode
  `\&12\catcode `\#12\catcode `\^12\catcode `\_12\catcode `\%12\relax}%
\providecommand \@@startlink[1]{}%
\providecommand \@@endlink[0]{}%
\providecommand \url  [0]{\begingroup\@sanitize@url \@url }%
\providecommand \@url [1]{\endgroup\@href {#1}{\urlprefix }}%
\providecommand \urlprefix  [0]{URL }%
\providecommand \Eprint [0]{\href }%
\providecommand \doibase [0]{http://dx.doi.org/}%
\providecommand \selectlanguage [0]{\@gobble}%
\providecommand \bibinfo  [0]{\@secondoftwo}%
\providecommand \bibfield  [0]{\@secondoftwo}%
\providecommand \translation [1]{[#1]}%
\providecommand \BibitemOpen [0]{}%
\providecommand \bibitemStop [0]{}%
\providecommand \bibitemNoStop [0]{.\EOS\space}%
\providecommand \EOS [0]{\spacefactor3000\relax}%
\providecommand \BibitemShut  [1]{\csname bibitem#1\endcsname}%
\let\auto@bib@innerbib\@empty
%</preamble>
\bibitem [{\citenamefont {Lee}\ \emph {et~al.}(2023{\natexlab{a}})\citenamefont
  {Lee}, \citenamefont {Kim}, \citenamefont {Im}, \citenamefont {An},
  \citenamefont {Kwon},\ and\ \citenamefont {Auh}}]{lee2023}%
  \BibitemOpen
  \bibfield  {author} {\bibinfo {author} {\bibfnamefont {S.}~\bibnamefont
  {Lee}}, \bibinfo {author} {\bibfnamefont {J.}~\bibnamefont {Kim}}, \bibinfo
  {author} {\bibfnamefont {S.}~\bibnamefont {Im}}, \bibinfo {author}
  {\bibfnamefont {S.}~\bibnamefont {An}}, \bibinfo {author} {\bibfnamefont
  {Y.-W.}\ \bibnamefont {Kwon}}, \ and\ \bibinfo {author} {\bibfnamefont
  {K.~H.}\ \bibnamefont {Auh}},\ }\href {\doibase 10.6111/JKCGCT.2023.33.2.061}
  {\bibfield  {journal} {\bibinfo  {journal} {Journal of the Korean Crystal
  Growth and Crystal Technology}\ }\textbf {\bibinfo {volume} {33}},\ \bibinfo
  {pages} {61} (\bibinfo {year} {2023}{\natexlab{a}})}\BibitemShut {NoStop}%
\bibitem [{\citenamefont {Lee}\ \emph {et~al.}(2023{\natexlab{b}})\citenamefont
  {Lee}, \citenamefont {Kim},\ and\ \citenamefont {Kwon}}]{lee2023a}%
  \BibitemOpen
  \bibfield  {author} {\bibinfo {author} {\bibfnamefont {S.}~\bibnamefont
  {Lee}}, \bibinfo {author} {\bibfnamefont {J.-H.}\ \bibnamefont {Kim}}, \ and\
  \bibinfo {author} {\bibfnamefont {Y.-W.}\ \bibnamefont {Kwon}},\ }\href
  {\doibase 10.48550/arXiv.2307.12008} {\enquote {\bibinfo {title} {The {{First
  Room-Temperature Ambient-Pressure Superconductor}}},}\ } (\bibinfo {year}
  {2023}{\natexlab{b}}),\ \Eprint {http://arxiv.org/abs/2307.12008}
  {arxiv:2307.12008 [cond-mat]} \BibitemShut {NoStop}%
\bibitem [{\citenamefont {Lee}\ \emph {et~al.}(2023{\natexlab{c}})\citenamefont
  {Lee}, \citenamefont {Kim}, \citenamefont {Kim}, \citenamefont {Im},
  \citenamefont {An},\ and\ \citenamefont {Auh}}]{lee2023b}%
  \BibitemOpen
  \bibfield  {author} {\bibinfo {author} {\bibfnamefont {S.}~\bibnamefont
  {Lee}}, \bibinfo {author} {\bibfnamefont {J.}~\bibnamefont {Kim}}, \bibinfo
  {author} {\bibfnamefont {H.-T.}\ \bibnamefont {Kim}}, \bibinfo {author}
  {\bibfnamefont {S.}~\bibnamefont {Im}}, \bibinfo {author} {\bibfnamefont
  {S.}~\bibnamefont {An}}, \ and\ \bibinfo {author} {\bibfnamefont {K.~H.}\
  \bibnamefont {Auh}},\ }\href {\doibase 10.48550/arXiv.2307.12037} {\enquote
  {\bibinfo {title} {Superconductor
  {{Pb}}$_{10-x}${{Cu}}$_x$({{PO}}$_4$)$_6${{O}} showing levitation at room
  temperature and atmospheric pressure and mechanism},}\ } (\bibinfo {year}
  {2023}{\natexlab{c}}),\ \Eprint {http://arxiv.org/abs/2307.12037}
  {arxiv:2307.12037 [cond-mat]} \BibitemShut {NoStop}%
\bibitem [{\citenamefont {Si}\ and\ \citenamefont {Held}(2023)}]{si2023}%
  \BibitemOpen
  \bibfield  {author} {\bibinfo {author} {\bibfnamefont {L.}~\bibnamefont
  {Si}}\ and\ \bibinfo {author} {\bibfnamefont {K.}~\bibnamefont {Held}},\
  }\href {\doibase 10.48550/arXiv.2308.00676} {\enquote {\bibinfo {title}
  {Electronic structure of the putative room-temperature superconductor
  {{Pb}}$_9${{Cu}}({{PO}}$_4$)$_6${{O}}},}\ } (\bibinfo {year} {2023}),\
  \Eprint {http://arxiv.org/abs/2308.00676} {arxiv:2308.00676 [cond-mat]}
  \BibitemShut {NoStop}%
\bibitem [{\citenamefont {Si}\ \emph {et~al.}(2023)\citenamefont {Si},
  \citenamefont {Wallerberger}, \citenamefont {Smolyanyuk}, \citenamefont {{di
  Cataldo}}, \citenamefont {Tomczak},\ and\ \citenamefont {Held}}]{Si2023b}%
  \BibitemOpen
  \bibfield  {author} {\bibinfo {author} {\bibfnamefont {L.}~\bibnamefont
  {Si}}, \bibinfo {author} {\bibfnamefont {M.}~\bibnamefont {Wallerberger}},
  \bibinfo {author} {\bibfnamefont {A.}~\bibnamefont {Smolyanyuk}}, \bibinfo
  {author} {\bibfnamefont {S.}~\bibnamefont {{di Cataldo}}}, \bibinfo {author}
  {\bibfnamefont {J.~M.}\ \bibnamefont {Tomczak}}, \ and\ \bibinfo {author}
  {\bibfnamefont {K.}~\bibnamefont {Held}},\ }\href {\doibase
  10.48550/arXiv.2308.04427} {\enquote {\bibinfo {title}
  {Pb$_{10-x}${{Cu}}$_x$({{PO}}$_4$)$_6${{O}}: A {{Mott}} or charge transfer
  insulator in need of further doping for (super)conductivity},}\ } (\bibinfo
  {year} {2023}),\ \Eprint {http://arxiv.org/abs/2308.04427} {arxiv:2308.04427
  [cond-mat]} \BibitemShut {NoStop}%
\bibitem [{\citenamefont {Korotin}\ \emph {et~al.}(2023)\citenamefont
  {Korotin}, \citenamefont {Novoselov}, \citenamefont {Shorikov}, \citenamefont
  {Anisimov},\ and\ \citenamefont {Oganov}}]{Korotin2023}%
  \BibitemOpen
  \bibfield  {author} {\bibinfo {author} {\bibfnamefont {D.~M.}\ \bibnamefont
  {Korotin}}, \bibinfo {author} {\bibfnamefont {D.~Y.}\ \bibnamefont
  {Novoselov}}, \bibinfo {author} {\bibfnamefont {A.~O.}\ \bibnamefont
  {Shorikov}}, \bibinfo {author} {\bibfnamefont {V.~I.}\ \bibnamefont
  {Anisimov}}, \ and\ \bibinfo {author} {\bibfnamefont {A.~R.}\ \bibnamefont
  {Oganov}},\ }\href {\doibase 10.48550/arXiv.2308.04301} {\enquote {\bibinfo
  {title} {Electronic correlations in promising room-temperature superconductor
  {{Pb}}$_9${{Cu}}({{PO}}$_4$)$_6${{O}}: A {{DFT}}+{{DMFT}} study},}\ }
  (\bibinfo {year} {2023}),\ \Eprint {http://arxiv.org/abs/2308.04301}
  {arxiv:2308.04301 [cond-mat]} \BibitemShut {NoStop}%
\bibitem [{\citenamefont {Yue}\ \emph {et~al.}(2023)\citenamefont {Yue},
  \citenamefont {Christiansson},\ and\ \citenamefont
  {Werner}}]{yue2023correlated}%
  \BibitemOpen
  \bibfield  {author} {\bibinfo {author} {\bibfnamefont {C.}~\bibnamefont
  {Yue}}, \bibinfo {author} {\bibfnamefont {V.}~\bibnamefont {Christiansson}},
  \ and\ \bibinfo {author} {\bibfnamefont {P.}~\bibnamefont {Werner}},\ }\href
  {\doibase 10.48550/arXiv.2308.04976} {\enquote {\bibinfo {title} {Correlated
  electronic structure of {{Pb}}$_{10-x}${{Cu}}$_x$({{PO4}})$_6${{O}}},}\ }
  (\bibinfo {year} {2023}),\ \Eprint {http://arxiv.org/abs/2308.04976}
  {arxiv:2308.04976 [cond-mat]} \BibitemShut {NoStop}%
\bibitem [{\citenamefont {Liu}\ \emph {et~al.}(2023{\natexlab{a}})\citenamefont
  {Liu}, \citenamefont {Yu}, \citenamefont {Li}, \citenamefont {Wang},
  \citenamefont {Lai}, \citenamefont {Sun}, \citenamefont {Chen},\ and\
  \citenamefont {Liu}}]{liu2023symmetry}%
  \BibitemOpen
  \bibfield  {author} {\bibinfo {author} {\bibfnamefont {J.}~\bibnamefont
  {Liu}}, \bibinfo {author} {\bibfnamefont {T.}~\bibnamefont {Yu}}, \bibinfo
  {author} {\bibfnamefont {J.}~\bibnamefont {Li}}, \bibinfo {author}
  {\bibfnamefont {J.}~\bibnamefont {Wang}}, \bibinfo {author} {\bibfnamefont
  {J.}~\bibnamefont {Lai}}, \bibinfo {author} {\bibfnamefont {Y.}~\bibnamefont
  {Sun}}, \bibinfo {author} {\bibfnamefont {X.-Q.}\ \bibnamefont {Chen}}, \
  and\ \bibinfo {author} {\bibfnamefont {P.}~\bibnamefont {Liu}},\ }\href
  {\doibase 10.48550/arXiv.2308.11766} {\enquote {\bibinfo {title} {Symmetry
  breaking induced insulating electronic state in
  {{Pb}}$_{9}${{Cu}}({{PO}}$_4$)$_6${{O}}},}\ } (\bibinfo {year}
  {2023}{\natexlab{a}}),\ \Eprint {http://arxiv.org/abs/2308.11766}
  {arxiv:2308.11766 [cond-mat]} \BibitemShut {NoStop}%
\bibitem [{\citenamefont {Georgescu}(2023)}]{georgescu2023}%
  \BibitemOpen
  \bibfield  {author} {\bibinfo {author} {\bibfnamefont {A.~B.}\ \bibnamefont
  {Georgescu}},\ }\href {\doibase 10.48550/arXiv.2308.07295} {\enquote
  {\bibinfo {title} {Cu-doped {{Pb}}$_{10}$({{PO}}$_4$)$_6${{O}}, and {{V}}
  doped {{SrTiO}}$_3$ -- a tutorial on electron-crystal lattice coupling in
  insulating materials with transition metal dopants},}\ } (\bibinfo {year}
  {2023}),\ \Eprint {http://arxiv.org/abs/2308.07295} {arxiv:2308.07295
  [cond-mat]} \BibitemShut {NoStop}%
\bibitem [{\citenamefont {Kumar}\ \emph {et~al.}(2023)\citenamefont {Kumar},
  \citenamefont {Karn}, \citenamefont {Kumar},\ and\ \citenamefont
  {Awana}}]{kumar2023absence}%
  \BibitemOpen
  \bibfield  {author} {\bibinfo {author} {\bibfnamefont {K.}~\bibnamefont
  {Kumar}}, \bibinfo {author} {\bibfnamefont {N.~K.}\ \bibnamefont {Karn}},
  \bibinfo {author} {\bibfnamefont {Y.}~\bibnamefont {Kumar}}, \ and\ \bibinfo
  {author} {\bibfnamefont {V.~P.~S.}\ \bibnamefont {Awana}},\ }\href {\doibase
  10.48550/arXiv.2308.03544} {\enquote {\bibinfo {title} {Absence of
  superconductivity in {{LK-99}} at ambient conditions},}\ } (\bibinfo {year}
  {2023}),\ \Eprint {http://arxiv.org/abs/2308.03544} {arxiv:2308.03544
  [cond-mat]} \BibitemShut {NoStop}%
\bibitem [{\citenamefont {Puphal}\ \emph {et~al.}(2023)\citenamefont {Puphal},
  \citenamefont {Akbar}, \citenamefont {Hepting}, \citenamefont {Goering},
  \citenamefont {Isobe}, \citenamefont {Nugroho},\ and\ \citenamefont
  {Keimer}}]{puphal2023single}%
  \BibitemOpen
  \bibfield  {author} {\bibinfo {author} {\bibfnamefont {P.}~\bibnamefont
  {Puphal}}, \bibinfo {author} {\bibfnamefont {M.~Y.~P.}\ \bibnamefont
  {Akbar}}, \bibinfo {author} {\bibfnamefont {M.}~\bibnamefont {Hepting}},
  \bibinfo {author} {\bibfnamefont {E.}~\bibnamefont {Goering}}, \bibinfo
  {author} {\bibfnamefont {M.}~\bibnamefont {Isobe}}, \bibinfo {author}
  {\bibfnamefont {A.~A.}\ \bibnamefont {Nugroho}}, \ and\ \bibinfo {author}
  {\bibfnamefont {B.}~\bibnamefont {Keimer}},\ }\href {\doibase
  10.48550/arXiv.2308.06256} {\enquote {\bibinfo {title} {Single crystal
  synthesis, structure, and magnetism of
  {{Pb}}$_{10-x}${{Cu}}$_x$({{PO}}$_4$)$_6${{O}}},}\ } (\bibinfo {year}
  {2023}),\ \Eprint {http://arxiv.org/abs/2308.06256} {2308.06256 [cond-mat]}
  \BibitemShut {NoStop}%
\bibitem [{\citenamefont {Jiang}\ \emph {et~al.}(2023)\citenamefont {Jiang},
  \citenamefont {Lee}, \citenamefont {{Herzog-Arbeitman}}, \citenamefont {Yu},
  \citenamefont {Feng}, \citenamefont {Hu}, \citenamefont {C{\u a}lug{\u a}ru},
  \citenamefont {Brodale}, \citenamefont {Gormley}, \citenamefont {Vergniory},
  \citenamefont {Felser}, \citenamefont {{Blanco-Canosa}}, \citenamefont
  {Hendon}, \citenamefont {Schoop},\ and\ \citenamefont
  {Bernevig}}]{jiang2023pb}%
  \BibitemOpen
  \bibfield  {author} {\bibinfo {author} {\bibfnamefont {Y.}~\bibnamefont
  {Jiang}}, \bibinfo {author} {\bibfnamefont {S.~B.}\ \bibnamefont {Lee}},
  \bibinfo {author} {\bibfnamefont {J.}~\bibnamefont {{Herzog-Arbeitman}}},
  \bibinfo {author} {\bibfnamefont {J.}~\bibnamefont {Yu}}, \bibinfo {author}
  {\bibfnamefont {X.}~\bibnamefont {Feng}}, \bibinfo {author} {\bibfnamefont
  {H.}~\bibnamefont {Hu}}, \bibinfo {author} {\bibfnamefont {D.}~\bibnamefont
  {C{\u a}lug{\u a}ru}}, \bibinfo {author} {\bibfnamefont {P.~S.}\ \bibnamefont
  {Brodale}}, \bibinfo {author} {\bibfnamefont {E.~L.}\ \bibnamefont
  {Gormley}}, \bibinfo {author} {\bibfnamefont {M.~G.}\ \bibnamefont
  {Vergniory}}, \bibinfo {author} {\bibfnamefont {C.}~\bibnamefont {Felser}},
  \bibinfo {author} {\bibfnamefont {S.}~\bibnamefont {{Blanco-Canosa}}},
  \bibinfo {author} {\bibfnamefont {C.~H.}\ \bibnamefont {Hendon}}, \bibinfo
  {author} {\bibfnamefont {L.~M.}\ \bibnamefont {Schoop}}, \ and\ \bibinfo
  {author} {\bibfnamefont {B.~A.}\ \bibnamefont {Bernevig}},\ }\href {\doibase
  10.48550/arXiv.2308.05143} {\enquote {\bibinfo {title}
  {Pb$_9${{Cu}}({{PO4}})$_6$({{OH}})$_2$: {{Phonon}} bands, {{Localized Flat
  Band Magnetism}}, {{Models}}, and {{Chemical Analysis}}},}\ } (\bibinfo
  {year} {2023}),\ \Eprint {http://arxiv.org/abs/2308.05143} {arxiv:2308.05143
  [cond-mat]} \BibitemShut {NoStop}%
\bibitem [{\citenamefont {Liu}\ \emph {et~al.}(2023{\natexlab{b}})\citenamefont
  {Liu}, \citenamefont {Cheng}, \citenamefont {Zhang}, \citenamefont {Xu},
  \citenamefont {Li}, \citenamefont {Shi}, \citenamefont {Yuan}, \citenamefont
  {Xu}, \citenamefont {Zhou}, \citenamefont {Zhu}, \citenamefont {Sun},
  \citenamefont {Wu}, \citenamefont {Luo}, \citenamefont {Jin},\ and\
  \citenamefont {Li}}]{liu2023}%
  \BibitemOpen
  \bibfield  {author} {\bibinfo {author} {\bibfnamefont {C.}~\bibnamefont
  {Liu}}, \bibinfo {author} {\bibfnamefont {W.}~\bibnamefont {Cheng}}, \bibinfo
  {author} {\bibfnamefont {X.}~\bibnamefont {Zhang}}, \bibinfo {author}
  {\bibfnamefont {J.}~\bibnamefont {Xu}}, \bibinfo {author} {\bibfnamefont
  {J.}~\bibnamefont {Li}}, \bibinfo {author} {\bibfnamefont {Q.}~\bibnamefont
  {Shi}}, \bibinfo {author} {\bibfnamefont {C.}~\bibnamefont {Yuan}}, \bibinfo
  {author} {\bibfnamefont {L.}~\bibnamefont {Xu}}, \bibinfo {author}
  {\bibfnamefont {H.}~\bibnamefont {Zhou}}, \bibinfo {author} {\bibfnamefont
  {S.}~\bibnamefont {Zhu}}, \bibinfo {author} {\bibfnamefont {J.}~\bibnamefont
  {Sun}}, \bibinfo {author} {\bibfnamefont {W.}~\bibnamefont {Wu}}, \bibinfo
  {author} {\bibfnamefont {J.}~\bibnamefont {Luo}}, \bibinfo {author}
  {\bibfnamefont {K.}~\bibnamefont {Jin}}, \ and\ \bibinfo {author}
  {\bibfnamefont {Y.}~\bibnamefont {Li}},\ }\href {\doibase
  10.1103/PhysRevMaterials.7.084804} {\bibfield  {journal} {\bibinfo  {journal}
  {Physical Review Materials}\ }\textbf {\bibinfo {volume} {7}},\ \bibinfo
  {pages} {084804} (\bibinfo {year} {2023}{\natexlab{b}})}\BibitemShut
  {NoStop}%
\bibitem [{\citenamefont {Wang}\ \emph {et~al.}(2023)\citenamefont {Wang},
  \citenamefont {Liu}, \citenamefont {Ge}, \citenamefont {Ji}, \citenamefont
  {Ji}, \citenamefont {Liu}, \citenamefont {Ai}, \citenamefont {Ma},
  \citenamefont {Qi},\ and\ \citenamefont {Wang}}]{wang2023ferromagnetic}%
  \BibitemOpen
  \bibfield  {author} {\bibinfo {author} {\bibfnamefont {P.}~\bibnamefont
  {Wang}}, \bibinfo {author} {\bibfnamefont {X.}~\bibnamefont {Liu}}, \bibinfo
  {author} {\bibfnamefont {J.}~\bibnamefont {Ge}}, \bibinfo {author}
  {\bibfnamefont {C.}~\bibnamefont {Ji}}, \bibinfo {author} {\bibfnamefont
  {H.}~\bibnamefont {Ji}}, \bibinfo {author} {\bibfnamefont {Y.}~\bibnamefont
  {Liu}}, \bibinfo {author} {\bibfnamefont {Y.}~\bibnamefont {Ai}}, \bibinfo
  {author} {\bibfnamefont {G.}~\bibnamefont {Ma}}, \bibinfo {author}
  {\bibfnamefont {S.}~\bibnamefont {Qi}}, \ and\ \bibinfo {author}
  {\bibfnamefont {J.}~\bibnamefont {Wang}},\ }\href {\doibase
  10.1007/s44214-023-00035-z} {\bibfield  {journal} {\bibinfo  {journal}
  {Quantum Frontiers}\ }\textbf {\bibinfo {volume} {2}},\ \bibinfo {pages} {1}
  (\bibinfo {year} {2023})}\BibitemShut {NoStop}%
\bibitem [{\citenamefont {Zhu}\ \emph {et~al.}(2023)\citenamefont {Zhu},
  \citenamefont {Wu}, \citenamefont {Li},\ and\ \citenamefont {Luo}}]{Zhu2023}%
  \BibitemOpen
  \bibfield  {author} {\bibinfo {author} {\bibfnamefont {S.}~\bibnamefont
  {Zhu}}, \bibinfo {author} {\bibfnamefont {W.}~\bibnamefont {Wu}}, \bibinfo
  {author} {\bibfnamefont {Z.}~\bibnamefont {Li}}, \ and\ \bibinfo {author}
  {\bibfnamefont {J.}~\bibnamefont {Luo}},\ }\href {\doibase
  10.48550/arXiv.2308.04353} {\enquote {\bibinfo {title} {First order
  transition in {{Pb}}$_{10-x}${{Cu}}$_x$({{PO}}$_4$)$_6${{O}} ($0.9<x<1.1$)
  containing {{Cu$_2$S}}},}\ } (\bibinfo {year} {2023}),\ \Eprint
  {http://arxiv.org/abs/2308.04353} {arxiv:2308.04353 [cond-mat]} \BibitemShut
  {NoStop}%
\bibitem [{\citenamefont {Jain}(2023)}]{Jain2023}%
  \BibitemOpen
  \bibfield  {author} {\bibinfo {author} {\bibfnamefont {P.~K.}\ \bibnamefont
  {Jain}},\ }\href {\doibase 10.1021/acs.jpcc.3c05684} {\bibfield  {journal}
  {\bibinfo  {journal} {The Journal of Physical Chemistry C}\ }\textbf
  {\bibinfo {volume} {127}},\ \bibinfo {pages} {18253} (\bibinfo {year}
  {2023})},\ \Eprint {http://arxiv.org/abs/2308.05222} {arxiv:2308.05222
  [cond-mat]} \BibitemShut {NoStop}%
\bibitem [{\citenamefont {Griffin}(2023)}]{griffin2023}%
  \BibitemOpen
  \bibfield  {author} {\bibinfo {author} {\bibfnamefont {S.~M.}\ \bibnamefont
  {Griffin}},\ }\href {\doibase 10.48550/arXiv.2307.16892} {\enquote {\bibinfo
  {title} {Origin of correlated isolated flat bands in copper-substituted lead
  phosphate apatite},}\ } (\bibinfo {year} {2023}),\ \Eprint
  {http://arxiv.org/abs/2307.16892} {arxiv:2307.16892 [cond-mat]} \BibitemShut
  {NoStop}%
\bibitem [{\citenamefont {Lai}\ \emph {et~al.}(2024)\citenamefont {Lai},
  \citenamefont {Li}, \citenamefont {Liu}, \citenamefont {Sun},\ and\
  \citenamefont {Chen}}]{lai2024}%
  \BibitemOpen
  \bibfield  {author} {\bibinfo {author} {\bibfnamefont {J.}~\bibnamefont
  {Lai}}, \bibinfo {author} {\bibfnamefont {J.}~\bibnamefont {Li}}, \bibinfo
  {author} {\bibfnamefont {P.}~\bibnamefont {Liu}}, \bibinfo {author}
  {\bibfnamefont {Y.}~\bibnamefont {Sun}}, \ and\ \bibinfo {author}
  {\bibfnamefont {X.-Q.}\ \bibnamefont {Chen}},\ }\href {\doibase
  10.1016/j.jmst.2023.08.001} {\bibfield  {journal} {\bibinfo  {journal}
  {Journal of Materials Science \& Technology}\ }\textbf {\bibinfo {volume}
  {171}},\ \bibinfo {pages} {66} (\bibinfo {year} {2024})}\BibitemShut
  {NoStop}%
\bibitem [{\citenamefont {{Cabezas-Escares}}\ \emph {et~al.}(2023)\citenamefont
  {{Cabezas-Escares}}, \citenamefont {Barrera}, \citenamefont {Lavroff},
  \citenamefont {Cardenas},\ and\ \citenamefont {Munoz}}]{cabezas-escares2023}%
  \BibitemOpen
  \bibfield  {author} {\bibinfo {author} {\bibfnamefont {J.}~\bibnamefont
  {{Cabezas-Escares}}}, \bibinfo {author} {\bibfnamefont {N.~F.}\ \bibnamefont
  {Barrera}}, \bibinfo {author} {\bibfnamefont {R.~H.}\ \bibnamefont
  {Lavroff}}, \bibinfo {author} {\bibfnamefont {A.~N. A.~C.}\ \bibnamefont
  {Cardenas}}, \ and\ \bibinfo {author} {\bibfnamefont {F.}~\bibnamefont
  {Munoz}},\ }\href {\doibase 10.48550/arXiv.2308.01135} {\enquote {\bibinfo
  {title} {Theoretical insight on the {{LK-99}} material ({{Large}} update)},}\
  } (\bibinfo {year} {2023}),\ \Eprint {http://arxiv.org/abs/2308.01135}
  {arxiv:2308.01135 [cond-mat]} \BibitemShut {NoStop}%
\bibitem [{\citenamefont {Zhou}\ and\ \citenamefont {Franz}(2023)}]{zhou2023}%
  \BibitemOpen
  \bibfield  {author} {\bibinfo {author} {\bibfnamefont {B.~T.}\ \bibnamefont
  {Zhou}}\ and\ \bibinfo {author} {\bibfnamefont {M.}~\bibnamefont {Franz}},\
  }\href {\doibase 10.48550/arXiv.2308.07408} {\enquote {\bibinfo {title}
  {Cu-substituted lead phosphate apatite as an inversion-asymmetric {{Weyl}}
  semimetal},}\ } (\bibinfo {year} {2023}),\ \Eprint
  {http://arxiv.org/abs/2308.07408} {arxiv:2308.07408 [cond-mat]} \BibitemShut
  {NoStop}%
\bibitem [{\citenamefont {Hirschmann}\ and\ \citenamefont
  {Mitscherling}(2023)}]{hirschmann2023}%
  \BibitemOpen
  \bibfield  {author} {\bibinfo {author} {\bibfnamefont {M.~M.}\ \bibnamefont
  {Hirschmann}}\ and\ \bibinfo {author} {\bibfnamefont {J.}~\bibnamefont
  {Mitscherling}},\ }\href {\doibase 10.48550/arXiv.2308.03751} {\enquote
  {\bibinfo {title} {Minimal model for double {{Weyl}} points, multiband
  quantum geometry, and singular flat band inspired by {{LK-99}}},}\ }
  (\bibinfo {year} {2023}),\ \Eprint {http://arxiv.org/abs/2308.03751}
  {arxiv:2308.03751 [cond-mat]} \BibitemShut {NoStop}%
\bibitem [{\citenamefont {Yang}\ \emph {et~al.}(2023)\citenamefont {Yang},
  \citenamefont {Liu},\ and\ \citenamefont {Zhong}}]{yang2023ab}%
  \BibitemOpen
  \bibfield  {author} {\bibinfo {author} {\bibfnamefont {S.}~\bibnamefont
  {Yang}}, \bibinfo {author} {\bibfnamefont {G.}~\bibnamefont {Liu}}, \ and\
  \bibinfo {author} {\bibfnamefont {Y.}~\bibnamefont {Zhong}},\ }\href
  {\doibase 10.48550/arXiv.2308.13938} {\enquote {\bibinfo {title} {Ab initio
  {{Investigations}} on the {{Electronic Properties}} and {{Stability}} of
  {{Cu-Substituted Lead Apatite}} ({{LK-99}}) family with different doping
  concentrations (x=0, 1, 2)},}\ } (\bibinfo {year} {2023}),\ \Eprint
  {http://arxiv.org/abs/2308.13938} {arxiv:2308.13938 [cond-mat]} \BibitemShut
  {NoStop}%
\bibitem [{\citenamefont {Koepernik}\ and\ \citenamefont
  {Eschrig}(1999)}]{koepernik1999}%
  \BibitemOpen
  \bibfield  {author} {\bibinfo {author} {\bibfnamefont {K.}~\bibnamefont
  {Koepernik}}\ and\ \bibinfo {author} {\bibfnamefont {H.}~\bibnamefont
  {Eschrig}},\ }\href {\doibase 10.1103/PhysRevB.59.1743} {\bibfield  {journal}
  {\bibinfo  {journal} {Physical Review B}\ }\textbf {\bibinfo {volume} {59}},\
  \bibinfo {pages} {1743} (\bibinfo {year} {1999})}\BibitemShut {NoStop}%
\bibitem [{\citenamefont {Perdew}\ and\ \citenamefont
  {Wang}(1992)}]{perdew1992}%
  \BibitemOpen
  \bibfield  {author} {\bibinfo {author} {\bibfnamefont {J.~P.}\ \bibnamefont
  {Perdew}}\ and\ \bibinfo {author} {\bibfnamefont {Y.}~\bibnamefont {Wang}},\
  }\href {\doibase 10.1103/PhysRevB.45.13244} {\bibfield  {journal} {\bibinfo
  {journal} {Physical Review B}\ }\textbf {\bibinfo {volume} {45}},\ \bibinfo
  {pages} {13244} (\bibinfo {year} {1992})}\BibitemShut {NoStop}%
\bibitem [{\citenamefont {Bauernfeind}\ \emph {et~al.}(2021)\citenamefont
  {Bauernfeind}, \citenamefont {Erhardt}, \citenamefont {Eck}, \citenamefont
  {Thakur}, \citenamefont {Gabel}, \citenamefont {Lee}, \citenamefont
  {Sch{\"a}fer}, \citenamefont {Moser}, \citenamefont {Di~Sante}, \citenamefont
  {Claessen},\ and\ \citenamefont {Sangiovanni}}]{bauernfeind2021}%
  \BibitemOpen
  \bibfield  {author} {\bibinfo {author} {\bibfnamefont {M.}~\bibnamefont
  {Bauernfeind}}, \bibinfo {author} {\bibfnamefont {J.}~\bibnamefont
  {Erhardt}}, \bibinfo {author} {\bibfnamefont {P.}~\bibnamefont {Eck}},
  \bibinfo {author} {\bibfnamefont {P.~K.}\ \bibnamefont {Thakur}}, \bibinfo
  {author} {\bibfnamefont {J.}~\bibnamefont {Gabel}}, \bibinfo {author}
  {\bibfnamefont {T.-L.}\ \bibnamefont {Lee}}, \bibinfo {author} {\bibfnamefont
  {J.}~\bibnamefont {Sch{\"a}fer}}, \bibinfo {author} {\bibfnamefont
  {S.}~\bibnamefont {Moser}}, \bibinfo {author} {\bibfnamefont
  {D.}~\bibnamefont {Di~Sante}}, \bibinfo {author} {\bibfnamefont
  {R.}~\bibnamefont {Claessen}}, \ and\ \bibinfo {author} {\bibfnamefont
  {G.}~\bibnamefont {Sangiovanni}},\ }\href {\doibase
  10.1038/s41467-021-25627-y} {\bibfield  {journal} {\bibinfo  {journal}
  {Nature Communications}\ }\textbf {\bibinfo {volume} {12}},\ \bibinfo {pages}
  {5396} (\bibinfo {year} {2021})}\BibitemShut {NoStop}%
\bibitem [{\citenamefont {Eck}\ \emph {et~al.}(2022)\citenamefont {Eck},
  \citenamefont {Ortix}, \citenamefont {Consiglio}, \citenamefont {Erhardt},
  \citenamefont {Bauernfeind}, \citenamefont {Moser}, \citenamefont {Claessen},
  \citenamefont {Di~Sante},\ and\ \citenamefont {Sangiovanni}}]{eck2022}%
  \BibitemOpen
  \bibfield  {author} {\bibinfo {author} {\bibfnamefont {P.}~\bibnamefont
  {Eck}}, \bibinfo {author} {\bibfnamefont {C.}~\bibnamefont {Ortix}}, \bibinfo
  {author} {\bibfnamefont {A.}~\bibnamefont {Consiglio}}, \bibinfo {author}
  {\bibfnamefont {J.}~\bibnamefont {Erhardt}}, \bibinfo {author} {\bibfnamefont
  {M.}~\bibnamefont {Bauernfeind}}, \bibinfo {author} {\bibfnamefont
  {S.}~\bibnamefont {Moser}}, \bibinfo {author} {\bibfnamefont
  {R.}~\bibnamefont {Claessen}}, \bibinfo {author} {\bibfnamefont
  {D.}~\bibnamefont {Di~Sante}}, \ and\ \bibinfo {author} {\bibfnamefont
  {G.}~\bibnamefont {Sangiovanni}},\ }\href {\doibase
  10.1103/PhysRevB.106.195143} {\bibfield  {journal} {\bibinfo  {journal}
  {Physical Review B}\ }\textbf {\bibinfo {volume} {106}},\ \bibinfo {pages}
  {195143} (\bibinfo {year} {2022})}\BibitemShut {NoStop}%
\bibitem [{\citenamefont {Tsirkin}\ \emph {et~al.}(2017)\citenamefont
  {Tsirkin}, \citenamefont {Souza},\ and\ \citenamefont
  {Vanderbilt}}]{tsirkin2017}%
  \BibitemOpen
  \bibfield  {author} {\bibinfo {author} {\bibfnamefont {S.~S.}\ \bibnamefont
  {Tsirkin}}, \bibinfo {author} {\bibfnamefont {I.}~\bibnamefont {Souza}}, \
  and\ \bibinfo {author} {\bibfnamefont {D.}~\bibnamefont {Vanderbilt}},\
  }\href {\doibase 10.1103/PhysRevB.96.045102} {\bibfield  {journal} {\bibinfo
  {journal} {Physical Review B}\ }\textbf {\bibinfo {volume} {96}},\ \bibinfo
  {pages} {045102} (\bibinfo {year} {2017})}\BibitemShut {NoStop}%
\bibitem [{\citenamefont {Nielsen}\ and\ \citenamefont
  {Ninomiya}(1981{\natexlab{a}})}]{nielsen1981}%
  \BibitemOpen
  \bibfield  {author} {\bibinfo {author} {\bibfnamefont {H.~B.}\ \bibnamefont
  {Nielsen}}\ and\ \bibinfo {author} {\bibfnamefont {M.}~\bibnamefont
  {Ninomiya}},\ }\href {\doibase 10.1016/0550-3213(81)90361-8} {\bibfield
  {journal} {\bibinfo  {journal} {Nuclear Physics B}\ }\textbf {\bibinfo
  {volume} {185}},\ \bibinfo {pages} {20} (\bibinfo {year}
  {1981}{\natexlab{a}})}\BibitemShut {NoStop}%
\bibitem [{\citenamefont {Nielsen}\ and\ \citenamefont
  {Ninomiya}(1981{\natexlab{b}})}]{nielsen1981a}%
  \BibitemOpen
  \bibfield  {author} {\bibinfo {author} {\bibfnamefont {H.~B.}\ \bibnamefont
  {Nielsen}}\ and\ \bibinfo {author} {\bibfnamefont {M.}~\bibnamefont
  {Ninomiya}},\ }\href {\doibase 10.1016/0550-3213(81)90524-1} {\bibfield
  {journal} {\bibinfo  {journal} {Nuclear Physics B}\ }\textbf {\bibinfo
  {volume} {193}},\ \bibinfo {pages} {173} (\bibinfo {year}
  {1981}{\natexlab{b}})}\BibitemShut {NoStop}%
\bibitem [{\citenamefont {Xu}\ \emph {et~al.}(2020)\citenamefont {Xu},
  \citenamefont {Zhang}, \citenamefont {Koepernik}, \citenamefont {Shi},
  \citenamefont {{van den Brink}}, \citenamefont {Felser},\ and\ \citenamefont
  {Sun}}]{xu2020}%
  \BibitemOpen
  \bibfield  {author} {\bibinfo {author} {\bibfnamefont {Q.}~\bibnamefont
  {Xu}}, \bibinfo {author} {\bibfnamefont {Y.}~\bibnamefont {Zhang}}, \bibinfo
  {author} {\bibfnamefont {K.}~\bibnamefont {Koepernik}}, \bibinfo {author}
  {\bibfnamefont {W.}~\bibnamefont {Shi}}, \bibinfo {author} {\bibfnamefont
  {J.}~\bibnamefont {{van den Brink}}}, \bibinfo {author} {\bibfnamefont
  {C.}~\bibnamefont {Felser}}, \ and\ \bibinfo {author} {\bibfnamefont
  {Y.}~\bibnamefont {Sun}},\ }\href {\doibase 10.1038/s41524-020-0301-1}
  {\bibfield  {journal} {\bibinfo  {journal} {npj Computational Materials}\
  }\textbf {\bibinfo {volume} {6}},\ \bibinfo {pages} {1} (\bibinfo {year}
  {2020})}\BibitemShut {NoStop}%
\bibitem [{\citenamefont {Saini}\ \emph {et~al.}(2022)\citenamefont {Saini},
  \citenamefont {Laurien}, \citenamefont {Blaha},\ and\ \citenamefont
  {Rubel}}]{saini2022a}%
  \BibitemOpen
  \bibfield  {author} {\bibinfo {author} {\bibfnamefont {H.}~\bibnamefont
  {Saini}}, \bibinfo {author} {\bibfnamefont {M.}~\bibnamefont {Laurien}},
  \bibinfo {author} {\bibfnamefont {P.}~\bibnamefont {Blaha}}, \ and\ \bibinfo
  {author} {\bibfnamefont {O.}~\bibnamefont {Rubel}},\ }\href {\doibase
  10.1016/j.cpc.2021.108147} {\bibfield  {journal} {\bibinfo  {journal}
  {Computer Physics Communications}\ }\textbf {\bibinfo {volume} {270}},\
  \bibinfo {pages} {108147} (\bibinfo {year} {2022})}\BibitemShut {NoStop}%
\bibitem [{\citenamefont {Wu}\ \emph {et~al.}(2018)\citenamefont {Wu},
  \citenamefont {Zhang}, \citenamefont {Song}, \citenamefont {Troyer},\ and\
  \citenamefont {Soluyanov}}]{wu2018}%
  \BibitemOpen
  \bibfield  {author} {\bibinfo {author} {\bibfnamefont {Q.}~\bibnamefont
  {Wu}}, \bibinfo {author} {\bibfnamefont {S.}~\bibnamefont {Zhang}}, \bibinfo
  {author} {\bibfnamefont {H.-F.}\ \bibnamefont {Song}}, \bibinfo {author}
  {\bibfnamefont {M.}~\bibnamefont {Troyer}}, \ and\ \bibinfo {author}
  {\bibfnamefont {A.~A.}\ \bibnamefont {Soluyanov}},\ }\href {\doibase
  10.1016/j.cpc.2017.09.033} {\bibfield  {journal} {\bibinfo  {journal}
  {Computer Physics Communications}\ }\textbf {\bibinfo {volume} {224}},\
  \bibinfo {pages} {405} (\bibinfo {year} {2018})}\BibitemShut {NoStop}%
\bibitem [{\citenamefont {Rackauckas}\ and\ \citenamefont
  {Nie}(2017)}]{rackauckas2017}%
  \BibitemOpen
  \bibfield  {author} {\bibinfo {author} {\bibfnamefont {C.}~\bibnamefont
  {Rackauckas}}\ and\ \bibinfo {author} {\bibfnamefont {Q.}~\bibnamefont
  {Nie}},\ }\href {\doibase 10.5334/jors.151} {\bibfield  {journal} {\bibinfo
  {journal} {Journal of Open Research Software}\ }\textbf {\bibinfo {volume}
  {5}},\ \bibinfo {pages} {15} (\bibinfo {year} {2017})}\BibitemShut {NoStop}%
\bibitem [{\citenamefont {Wan}\ \emph {et~al.}(2011)\citenamefont {Wan},
  \citenamefont {Turner}, \citenamefont {Vishwanath},\ and\ \citenamefont
  {Savrasov}}]{wan2011}%
  \BibitemOpen
  \bibfield  {author} {\bibinfo {author} {\bibfnamefont {X.}~\bibnamefont
  {Wan}}, \bibinfo {author} {\bibfnamefont {A.~M.}\ \bibnamefont {Turner}},
  \bibinfo {author} {\bibfnamefont {A.}~\bibnamefont {Vishwanath}}, \ and\
  \bibinfo {author} {\bibfnamefont {S.~Y.}\ \bibnamefont {Savrasov}},\ }\href
  {\doibase 10.1103/PhysRevB.83.205101} {\bibfield  {journal} {\bibinfo
  {journal} {Phys. Rev. B}\ }\textbf {\bibinfo {volume} {83}},\ \bibinfo
  {pages} {205101} (\bibinfo {year} {2011})}\BibitemShut {NoStop}%
\bibitem [{\citenamefont {Belopolski}\ \emph {et~al.}(2016)\citenamefont
  {Belopolski}, \citenamefont {Xu}, \citenamefont {Sanchez}, \citenamefont
  {Chang}, \citenamefont {Guo}, \citenamefont {Neupane}, \citenamefont {Zheng},
  \citenamefont {Lee}, \citenamefont {Huang}, \citenamefont {Bian},
  \citenamefont {Alidoust}, \citenamefont {Chang}, \citenamefont {Wang},
  \citenamefont {Zhang}, \citenamefont {Bansil}, \citenamefont {Jeng},
  \citenamefont {Lin}, \citenamefont {Jia},\ and\ \citenamefont
  {Hasan}}]{belopolski2016}%
  \BibitemOpen
  \bibfield  {author} {\bibinfo {author} {\bibfnamefont {I.}~\bibnamefont
  {Belopolski}}, \bibinfo {author} {\bibfnamefont {S.-Y.}\ \bibnamefont {Xu}},
  \bibinfo {author} {\bibfnamefont {D.~S.}\ \bibnamefont {Sanchez}}, \bibinfo
  {author} {\bibfnamefont {G.}~\bibnamefont {Chang}}, \bibinfo {author}
  {\bibfnamefont {C.}~\bibnamefont {Guo}}, \bibinfo {author} {\bibfnamefont
  {M.}~\bibnamefont {Neupane}}, \bibinfo {author} {\bibfnamefont
  {H.}~\bibnamefont {Zheng}}, \bibinfo {author} {\bibfnamefont {C.-C.}\
  \bibnamefont {Lee}}, \bibinfo {author} {\bibfnamefont {S.-M.}\ \bibnamefont
  {Huang}}, \bibinfo {author} {\bibfnamefont {G.}~\bibnamefont {Bian}},
  \bibinfo {author} {\bibfnamefont {N.}~\bibnamefont {Alidoust}}, \bibinfo
  {author} {\bibfnamefont {T.-R.}\ \bibnamefont {Chang}}, \bibinfo {author}
  {\bibfnamefont {B.}~\bibnamefont {Wang}}, \bibinfo {author} {\bibfnamefont
  {X.}~\bibnamefont {Zhang}}, \bibinfo {author} {\bibfnamefont
  {A.}~\bibnamefont {Bansil}}, \bibinfo {author} {\bibfnamefont {H.-T.}\
  \bibnamefont {Jeng}}, \bibinfo {author} {\bibfnamefont {H.}~\bibnamefont
  {Lin}}, \bibinfo {author} {\bibfnamefont {S.}~\bibnamefont {Jia}}, \ and\
  \bibinfo {author} {\bibfnamefont {M.~Z.}\ \bibnamefont {Hasan}},\ }\href
  {\doibase 10.1103/PhysRevLett.116.066802} {\bibfield  {journal} {\bibinfo
  {journal} {Phys. Rev. Lett.}\ }\textbf {\bibinfo {volume} {116}},\ \bibinfo
  {pages} {066802} (\bibinfo {year} {2016})}\BibitemShut {NoStop}%
\bibitem [{\citenamefont {Armitage}\ \emph {et~al.}(2018)\citenamefont
  {Armitage}, \citenamefont {Mele},\ and\ \citenamefont
  {Vishwanath}}]{armitage2018}%
  \BibitemOpen
  \bibfield  {author} {\bibinfo {author} {\bibfnamefont {N.~P.}\ \bibnamefont
  {Armitage}}, \bibinfo {author} {\bibfnamefont {E.~J.}\ \bibnamefont {Mele}},
  \ and\ \bibinfo {author} {\bibfnamefont {A.}~\bibnamefont {Vishwanath}},\
  }\href {\doibase 10.1103/RevModPhys.90.015001} {\bibfield  {journal}
  {\bibinfo  {journal} {Reviews of Modern Physics}\ }\textbf {\bibinfo {volume}
  {90}},\ \bibinfo {pages} {015001} (\bibinfo {year} {2018})}\BibitemShut
  {NoStop}%
\bibitem [{\citenamefont {Haldane}(2014)}]{haldane2014}%
  \BibitemOpen
  \bibfield  {author} {\bibinfo {author} {\bibfnamefont {F.~D.~M.}\
  \bibnamefont {Haldane}},\ }\href {\doibase 10.48550/arXiv.1401.0529}
  {\enquote {\bibinfo {title} {Attachment of {{Surface}} "{{Fermi Arcs}}" to
  the {{Bulk Fermi Surface}}: "{{Fermi-Level Plumbing}}" in {{Topological
  Metals}}},}\ } (\bibinfo {year} {2014}),\ \Eprint
  {http://arxiv.org/abs/1401.0529} {arxiv:1401.0529 [cond-mat]} \BibitemShut
  {NoStop}%
\bibitem [{\citenamefont {Hoesch}\ \emph {et~al.}(2002)\citenamefont {Hoesch},
  \citenamefont {Greber}, \citenamefont {Petrov}, \citenamefont {Muntwiler},
  \citenamefont {Hengsberger}, \citenamefont {Auw{\"a}rter},\ and\
  \citenamefont {Osterwalder}}]{hoesch2002}%
  \BibitemOpen
  \bibfield  {author} {\bibinfo {author} {\bibfnamefont {M.}~\bibnamefont
  {Hoesch}}, \bibinfo {author} {\bibfnamefont {T.}~\bibnamefont {Greber}},
  \bibinfo {author} {\bibfnamefont {V.~N.}\ \bibnamefont {Petrov}}, \bibinfo
  {author} {\bibfnamefont {M.}~\bibnamefont {Muntwiler}}, \bibinfo {author}
  {\bibfnamefont {M.}~\bibnamefont {Hengsberger}}, \bibinfo {author}
  {\bibfnamefont {W.}~\bibnamefont {Auw{\"a}rter}}, \ and\ \bibinfo {author}
  {\bibfnamefont {J.}~\bibnamefont {Osterwalder}},\ }\href {\doibase
  10.1016/S0368-2048(02)00058-0} {\bibfield  {journal} {\bibinfo  {journal}
  {Journal of Electron Spectroscopy and Related Phenomena}\ }\bibinfo {series}
  {Frontiers in Photoemission Spectroscopy of Solids and Surfaces},\ \textbf
  {\bibinfo {volume} {124}},\ \bibinfo {pages} {263} (\bibinfo {year}
  {2002})}\BibitemShut {NoStop}%
\bibitem [{\citenamefont {Lv}\ \emph {et~al.}(2019)\citenamefont {Lv},
  \citenamefont {Qian},\ and\ \citenamefont {Ding}}]{lv2019}%
  \BibitemOpen
  \bibfield  {author} {\bibinfo {author} {\bibfnamefont {B.}~\bibnamefont
  {Lv}}, \bibinfo {author} {\bibfnamefont {T.}~\bibnamefont {Qian}}, \ and\
  \bibinfo {author} {\bibfnamefont {H.}~\bibnamefont {Ding}},\ }\href {\doibase
  10.1038/s42254-019-0088-5} {\bibfield  {journal} {\bibinfo  {journal} {Nature
  Reviews Physics}\ }\textbf {\bibinfo {volume} {1}},\ \bibinfo {pages} {609}
  (\bibinfo {year} {2019})}\BibitemShut {NoStop}%
\bibitem [{\citenamefont {Huang}\ \emph {et~al.}(2015)\citenamefont {Huang},
  \citenamefont {Xu}, \citenamefont {Belopolski}, \citenamefont {Lee},
  \citenamefont {Chang}, \citenamefont {Wang}, \citenamefont {Alidoust},
  \citenamefont {Bian}, \citenamefont {Neupane}, \citenamefont {Zhang},
  \citenamefont {Jia}, \citenamefont {Bansil}, \citenamefont {Lin},\ and\
  \citenamefont {Hasan}}]{huang2015}%
  \BibitemOpen
  \bibfield  {author} {\bibinfo {author} {\bibfnamefont {S.-M.}\ \bibnamefont
  {Huang}}, \bibinfo {author} {\bibfnamefont {S.-Y.}\ \bibnamefont {Xu}},
  \bibinfo {author} {\bibfnamefont {I.}~\bibnamefont {Belopolski}}, \bibinfo
  {author} {\bibfnamefont {C.-C.}\ \bibnamefont {Lee}}, \bibinfo {author}
  {\bibfnamefont {G.}~\bibnamefont {Chang}}, \bibinfo {author} {\bibfnamefont
  {B.}~\bibnamefont {Wang}}, \bibinfo {author} {\bibfnamefont {N.}~\bibnamefont
  {Alidoust}}, \bibinfo {author} {\bibfnamefont {G.}~\bibnamefont {Bian}},
  \bibinfo {author} {\bibfnamefont {M.}~\bibnamefont {Neupane}}, \bibinfo
  {author} {\bibfnamefont {C.}~\bibnamefont {Zhang}}, \bibinfo {author}
  {\bibfnamefont {S.}~\bibnamefont {Jia}}, \bibinfo {author} {\bibfnamefont
  {A.}~\bibnamefont {Bansil}}, \bibinfo {author} {\bibfnamefont
  {H.}~\bibnamefont {Lin}}, \ and\ \bibinfo {author} {\bibfnamefont {M.~Z.}\
  \bibnamefont {Hasan}},\ }\href {\doibase 10.1038/ncomms8373} {\bibfield
  {journal} {\bibinfo  {journal} {Nature Communications}\ }\textbf {\bibinfo
  {volume} {6}},\ \bibinfo {pages} {7373} (\bibinfo {year} {2015})}\BibitemShut
  {NoStop}%
\bibitem [{\citenamefont {Devizorova}\ and\ \citenamefont
  {Volkov}(2017)}]{devizorova2017}%
  \BibitemOpen
  \bibfield  {author} {\bibinfo {author} {\bibfnamefont {Z.~A.}\ \bibnamefont
  {Devizorova}}\ and\ \bibinfo {author} {\bibfnamefont {V.~A.}\ \bibnamefont
  {Volkov}},\ }\href {\doibase 10.1103/PhysRevB.95.081302} {\bibfield
  {journal} {\bibinfo  {journal} {Phys. Rev. B}\ }\textbf {\bibinfo {volume}
  {95}},\ \bibinfo {pages} {081302} (\bibinfo {year} {2017})}\BibitemShut
  {NoStop}%
\bibitem [{\citenamefont {McCormick}\ \emph {et~al.}(2017)\citenamefont
  {McCormick}, \citenamefont {Kimchi},\ and\ \citenamefont
  {Trivedi}}]{mccormick2017}%
  \BibitemOpen
  \bibfield  {author} {\bibinfo {author} {\bibfnamefont {T.~M.}\ \bibnamefont
  {McCormick}}, \bibinfo {author} {\bibfnamefont {I.}~\bibnamefont {Kimchi}}, \
  and\ \bibinfo {author} {\bibfnamefont {N.}~\bibnamefont {Trivedi}},\ }\href
  {\doibase 10.1103/PhysRevB.95.075133} {\bibfield  {journal} {\bibinfo
  {journal} {Physical Review B}\ }\textbf {\bibinfo {volume} {95}},\ \bibinfo
  {pages} {075133} (\bibinfo {year} {2017})}\BibitemShut {NoStop}%
\bibitem [{\citenamefont {Wang}\ \emph {et~al.}(2020)\citenamefont {Wang},
  \citenamefont {Lin}, \citenamefont {Jiang}, \citenamefont {Guo},\ and\
  \citenamefont {Jiang}}]{wang2020}%
  \BibitemOpen
  \bibfield  {author} {\bibinfo {author} {\bibfnamefont {H.-X.}\ \bibnamefont
  {Wang}}, \bibinfo {author} {\bibfnamefont {Z.-K.}\ \bibnamefont {Lin}},
  \bibinfo {author} {\bibfnamefont {B.}~\bibnamefont {Jiang}}, \bibinfo
  {author} {\bibfnamefont {G.-Y.}\ \bibnamefont {Guo}}, \ and\ \bibinfo
  {author} {\bibfnamefont {J.-H.}\ \bibnamefont {Jiang}},\ }\href {\doibase
  10.1103/PhysRevLett.125.146401} {\bibfield  {journal} {\bibinfo  {journal}
  {Physical Review Letters}\ }\textbf {\bibinfo {volume} {125}},\ \bibinfo
  {pages} {146401} (\bibinfo {year} {2020})}\BibitemShut {NoStop}%
\end{thebibliography}%

\end{document}